\documentclass[11pt]{amsart}
\usepackage{appendix}
\usepackage[nofiglist]{endfloat}
\usepackage{graphicx}
\usepackage{rotating}
\usepackage{pdflscape}
\usepackage[center]{caption}

\RequirePackage{amssymb}
\RequirePackage{enumerate}
\RequirePackage{amsfonts}
\RequirePackage{amsmath}
\RequirePackage{mathrsfs}
\allowdisplaybreaks[1]

\newcounter{fig}
\newcommand{\mybic}{\author{Gianluca Cassese}
                     \address{Universit\`{a} Milano Bicocca}
                     \email{gianluca.cassese@unimib.it}
                     \curraddr{Department of Economics, Statistics and Management 
                                  Building U7, Room 2097, via Bicocca 
                                  degli Arcimboldi 8, 20126 Milano - Italy}}

\newtheorem{theorem}{Theorem}
\theoremstyle{plain}

\newtheorem{assumption}{Assumption}

\newtheorem{lemma}{Lemma}

\newcommand{\F}{\mathscr{F}}
\newcommand{\R}{\mathbb{R}} 
\newcommand{\N}{\mathbb{N}}

\newcommand{\Bor}{\mathscr{B}}

\newcommand{\quot}[1]{\textit{``#1''}}

\newcommand{\abs}[1]{\vert #1\vert} 
\newcommand{\babs}[1]{\big\vert #1\big\vert}

\newcommand{\Der}[2]{d #1\left/d #2\right.}
\newcommand{\dDer}[2]{\dfrac{d #1}{d #2}}

\newcommand{\PD}[2]{\partial #1\left/\partial #2\right.}

\newcommand{\bnorm}[1]{\big\Vert #1\big\Vert}

\newcommand{\set}[1]{\mathbf{1}_{#1}}
\newcommand{\sset}[1]{\mathbf{1}_{\{#1\}}}

\newcommand {\ess}{\mathrm{ess}}
\begin{document} 
\title[Non Parametric Option Prices]{Non Parametric Estimates of Option Prices Using Superhedging}
\mybic 
\date\today 
\subjclass{G12, C14.} 
\keywords{
Bid/Ask spreads, Implied risk-neutral measure, Non parametric regression.
}

\thanks{
In writing the empirical applications of this paper, I benefited from conversations 
with Patrick Gagliardini. I am also grateful to Alexandru Popescu for helping me 
with the MDR-CBOE Dataset. All errors are my responsibility.
}

\begin{abstract} 
We propose a new non parametric technique to estimate the CALL function
based on the superhedging principle. Our approach does not require absence
of arbitrage and easily accommodates bid/ask spreads and other market 
imperfections. We prove some optimal statistical properties of our estimates.
As an application we first test the methodology on a simulated sample of
option prices and then on the S\&P 500 index options.
\end{abstract}

\maketitle

\section{Introduction} 

A classical exercise in the econometric analysis of financial markets is to estimate
option prices and the risk neutral probability (or its density) which is implicit in them, 
as suggested by the famous works of Breeden and Litzenberger \cite{breeden 
litzenberger} and of Banz and Miller \cite{banz miller}. Of special importance in this 
exercise is the use of non parametric techniques which have become popular in 
the last decades. Regardless of the econometric approach taken, a first step of 
crucial importance is purging the data from observations violating some no-arbitrage 
condition and thus conflicting with the conclusion that prices are the expected value 
of the asset discounted payoff computed with respect to the risk neutral probability, 
as assumed by Breeden and Litzenberger.

Recalcitrant observations may however originate from different sources other 
than  just pure mispricing. There are first some microstructural issues, such 
as the bid/ask spread or other transaction costs, which are not considered in 
the risk neutral approach. The common practice is to assume that these 
components are uncorrelated with the fundamental value of the assets and 
may therefore be disposed of with simple transformations, such as computing 
mid prices. Nevertheless, a more accurate investigation reveals that market 
makers often adjust the spread in response to pressure originating either from 
demand or supply, a fact that contributes to make the spread asymmetric and 
to induce a possibly significant correlation with the fundamental price. A second 
factor influencing market data is the existence of restrictions to trading, these 
too often overlooked in asset pricing models which typically assume the space of 
marketed claims to be linear. The extreme versions of these restrictions take 
the form of short selling prohibitions but even with less drastic market rules, 
the possibility of shorting assets is clearly delicate and subject to constraints 
such as the provision of appropriate margins the effect of which receives in 
general little attention -- if any -- in theoretical and applied work on options. In 
fact, violations of the lower bound for CALL options, a mispricing that may be 
exploited by shorting the underlying, are often more frequent than others 
associated with strategies involving positions which are easier to take. Some 
additional noise arises with strategies which prescribe to invest simultaneously 
on different markets, a characteristic which results not only in relevant fixed 
costs but also in the lack of trade synchronism thus inducing additional risks in the 
execution of arbitrage strategies. A further issue arises eventually in connection 
with the riskless asset appearing in virtually all financial models and definitely in 
all empirical exercises. The assumption of a riskless asset and its explicit 
identification in applied work appears more and more counter factual as the interbank 
market turmoil in the years 2007-08 and the sovereign debt crisis following it have 
clearly demonstrated. Not only during periods of crisis it cannot be assumed that 
bonds are riskless but it is also hard to assert that the implicit risks are uncorrelated 
with the equity market. 

Of course, the relevance of the preceding remarks would be much less disturbing 
if one could restrict attention to just the options market and consider only trading
strategies which involve long positions. Unfortunately in this restricted perspective
the fundamental premise of the Breeden Litzenberger exercise, i.e. that option prices 
are set on the basis of a risk neutral probability, cannot be assumed. Nevertheless,
even if real option prices may fail to possess some fundamental properties, the
superhedging price, conveniently computed will possibly satisfy them under mild
conditions which involve only very simple and realistic trading strategies. The first 
step of our econometric approach consists of replacing original option prices with
superhedging prices.

As a matter of fact this idea is not new. A\"it-Sahalia and Duarte \cite{ait sahalia duarte}
have inaugurated a two stage approach to non parametric estimation of option prices 
under shape restrictions which, in the first stage, prescribes to project market prices
following a methodology adapted from Dykstra \cite{dykstra}. This procedure enforces 
the shape restrictions required. In the second stage, the prices so transformed are 
smoothed out using a local polynomial technique, a methodology which generalizes 
suitably the kernel by replacing constant functions with polynomials of arbitrary but 
preassigned degree. It turns out, as will be shown in the body of the paper, that the 
projection technique suggested by Dykstra produces just the options superhedging price, 
$q^0(k)$ -- with $k$ the strike price. The need for additional smoothing arises because 
the CALL function obtained, $k\to q^0(k)$ is piecewise linear and therefore not 
informative enough for many a purpose, including the project to extract a probability 
density.

In a recent paper, \cite{ThOpt}, we have obtained a result which is particularly 
useful for the present purposes and that lays the ground for our non parametric 
procedure. It states, roughly put, that it is possible to compute explicitly the 
superhedging price $q^G(t)$ for a large class $G$ of derivatives written on the same 
underlying and with payoff depending in a convex and decreasing way on a positive 
parameter, $t$. Not only, but for each such family $G$ of derivatives it is possible to
extract from the prices $q^G(t)$ a probability measure, $\nu^G$, implicit in them 
in much the same way as suggested by Breeden and Litzenberger. A special case 
is of course the family $G^0$ corresponding to plain options but in general the choice
of the family $G$ is open. Our suggestion is to consider new derivatives with a payoff 
which, while being perfectly smooth, approximates the option payoff uniformly. Moreover, 
the degree of approximation should be made sample dependent so to obtain good 
asymptotic properties. As long as these properties are guaranteed, any candidate $G$ 
is suitable, more or less as the functional form of kernels is relatively unimportant in 
comparison with the choice of the bandwidth. We have found it easy to work with 
splines but it should made clear from the outset that our use of splines has nothing 
to do with the econometric approaches based on this class of functions, such as those 
reviewed by Eubank \cite{eubank} and successfully applied to options by Fengler 
and Hin \cite{fengler hin}. Our problem in fact is not that of smoothing the option 
\textit{prices} (or their implied volatility surface) but rather the option \textit{payoff}. 
Splines simply turn out to be a conveniently tractable tool from a computational point 
of view; in addition, the smoothness/goodness of fit trade off may be tuned conveniently 
via a parameter acting as the bandwidth in kernel estimation. Other smoothing
techniques may be employed, e.g. those based on some known probability density
function. As a term of comparison, we shall briefly discuss the result obtained by
smoothing options via the normal density rather than splines.

The econometric analysis of option prices has grown over the years to become 
almost a field of its own and has been masterly reviewed by Garcia, Ghysels and 
Renault in \cite{garcia ghysels renault}. The interested reader may find in their
work an exhaustive list of references that we shall not try to improve upon here, 
for reasons of brevity. The non parametric approach has itself produced quite a 
number of important contributions that are worth discussing briefly with no 
pretension of completeness%
\footnote{
A nice review of non parametric methods and their relevance for economics
is in Yatchew \cite{yatchew}.
}. 
A forerunner of this 
stream of studies is the paper by Jackwerth and Rubinstein \cite{jackwerth rubinstein} 
in which the parameters of a binomial tree are set so as to minimize the distance 
between binomial and actual option prices given a penalty for deviations from an initial 
\textit{a priori} distribution. The assumption of absence of arbitrage opportunities is 
absolutely crucial here and is embodied in the tree parameters. Estimating the option 
price produces simultaneously an estimate of the risk neutral measure. A\"it-Sahalia 
and Lo \cite{ait sahalia lo} estimate  the CALL function using the Nadaraya-Watson 
kernel and recover the risk-neutral density by computing the second derivative. This 
methodology has optimal asymptotic properties but relatively poor performance in 
small samples, a fact common to many non parametric methods and of special 
concern for options market. The curse of dimensionality problem, see 
\cite[p. 675]{yatchew}, requires to limit as much as possible the number of state 
variables. For example, the risk neutral density estimated in \cite{ait sahalia lo} via 
kernel smoothing is surely convergent to the true density but may fail even to be 
non negative in small samples \cite[footnote 11, p. 508]{ait sahalia lo}. To circumvent 
this problem the authors propose a semi nonparametric approach in which the price 
is computed according to the Black and Scholes formula in which the volatility 
function is estimated non parametrically. This choice has the clear advantage of 
guaranteeing the correct shape  of the CALL function. Shape restrictions are easily 
accommodated in parametric modeling but are much more troublesome in the non 
parametric approach. Papers implementing the nonparametric methodology with 
shape restrictions are less numerous and more recent. Further to A\"it-Sahalia 
and Duarte \cite{ait sahalia duarte}, another example of non parametric techniques 
incorporating shape restrictions is the paper by Yatchew and H\"ardle 
\cite{yatchew hardle} who follow a least squares approach in Sobolev spaces. 
There exist eventually several papers who adopt spline techniques to estimate
option prices and one should mention Fengler \cite{fengler}, Fengler and Hin 
\cite{fengler hin} and Yin, Wang and Qi \cite{yin wang qi}. 

We think that if our approach has any merit compared to the above references,
then this lies in its clear financial interpretation. In fact we obtain the smoothness 
of the CALL function by pricing an appropriate derivative rather than performing 
some local averaging or implementing some other statistical technique%
\footnote{
A partial exception is the XMM methodology proposed by Gagliardini et al. 
\cite{gagliardini gourieroux renault}, in which the GMM is applied with the 
additional constraint of reproducing a subset of given prices which are 
considered \textit{a priori} to be correct.
}.
Moreover,
our method allows to take selection effects into full account, a fact not always
clearly considered. Eventually the estimates produced are extremely tractable
computationally speaking and have desirable convergence properties. 

The paper is structured as follows. In section \ref{sec option} we introduce
the fundamental results obtained in \cite{ThOpt} and needed in our exercise,
together with the necessary notation. In section \ref{sec estimate} we illustrate
all details of our spline based technique and prove some of its properties. We 
also comment on the possibility of using given density functions. In section 
\ref{sec simulation} we perform some numerical experiments on option
prices generated by an a priori model. In particular we construct non
parametric interval estimates in the presence of a realistic structure of noise.
Eventually in section \ref{sec market} we apply the proposed methodology
to a sample of S\&P option prices. All proofs are in the Appendix.

Let us mention in closing that, although as dictated by the literature on this topic 
we start assuming the existence of a given probability space, $(\Omega,\F,P)$, 
the results developed in \cite{ThOpt} do not require this classical premise.

\section{Option Pricing}\label{sec option}
Most of this section is adapted from \cite{ThOpt}. Let the random variable $X$ 
represent hereafter the payoff of a given underlying and $K(X)$ the set of strike 
prices (including $k=0$) of all CALL options written on it. We assume $P(X>0)=1$
and $P(X<k)>0$ for all $k\in K(X)\setminus\{0\}$%
\footnote{
This corresponds to assuming that no quoted option is known to expire in the 
money with probability 1.
}. 
The portfolio consisting of one option with strike price $k\in K(X)$ and its price will 
be indicated by the symbols $\theta(k)$ and $q(k)$ respectively. Moreover, its payoff 
will be denoted with the symbol $g^0_k(X)=(X-k)^+$. As we only consider long 
positions, $q(k)$ will actually be the \textit{ask} price and we shall assume that it 
is positive. The set $\Theta$ of admissible trading strategies is a convex set of 
portfolios formed by taking long positions in the set of traded options. The price 
and the payoff of $\theta=\sum_{n=1}^N\alpha_n\theta(k_n)\in\Theta$ -- so that 
$\alpha_1,\ldots,\alpha_N\ge0$ and $k_1,\ldots,k_N\in K(X)$ -- are denoted 
respectively by
\begin{equation}
\label{q}
q(\theta)
	=
\sum_{n=1}^N\alpha_nq(k_n)
\qquad\text{and}\qquad
g^0_\theta(X)
	=
\sum_{n=1}^N\alpha_ng^0_{k_n}(X)
\end{equation}

The price function $q$ does not include possible fixed trading costs. The apparent linear
structure implicit in \eqref{q}, notwithstanding the existence of bid/ask spreads
(which many authors associate with subadditivity of prices, see \cite{ThOpt}),
is due to the restriction included in $\Theta$ that only long positions may be
assumed and the traditional anonymity of option trading. OTC trading of
options follows, as is well known, different pricing schemes, often ad hoc,
and our theory does not apply to these transactions%
\footnote{
More details on the assumptions behind the present construction are found in
\cite{ThOpt}.
}.

We define the superhedging price of an arbitrary random quantity $h:\Omega\to\R$
as%
\footnote{
In \cite{ThOpt} the functional $\pi$ is defined in much greater generality and
no probability measure $P$ is taken as exogenously given.
}

\begin{equation}
\label{pi}
\pi(h)
	=
\inf\Big\{\lambda q(\theta):
P\big(\lambda g^0_\theta(X)/(X\wedge1)\ge h\big)=1, 
\lambda>0,
\theta\in\Theta\Big\}
\end{equation}
Write $q^0(k)=\pi\big(g_k^0(X)/(X\wedge1)\big)$.
We say that an option with strike price $k$ is priced efficiently if 
$
q(k)=q^0(k)
$
and denote by $K^0(X)=\{j_0,\ldots,j_I\}$ the set of strike prices for which 
we have an efficient price. It is easily shown that necessarily $j_0=0$. Remark 
that the notion of efficiency adopted here is particularly poor as it only involves 
investments with long positions in CALL options and the underlying. Neither PUT 
options, Futures nor bonds are contemplated. This is desirable since the larger 
the set of derivatives involved the more likely is it that efficiency may fail, as in 
the case of the PUT/CALL parity.

We make the following

\begin{assumption}
\label{ass X}
Let $\bar X=\ess\sup X$ and assume that 
\begin{equation}
\bar X>j_I
\quad\text{and}\quad
\ess\sup(X\wedge z)
	=
\bar X\wedge z
\quad\text{ for each } z\in\R
\end{equation}
\end{assumption}

Denote by $\Gamma$ the set of convex functions $g:\R_+\to\R_+$ with 
$g(0)=0$ and $\lim_{n\to\infty}g(n)/n<\infty$. Clearly, the function $g^0_t$ 
introduced above and corresponding to the option payoff is an element of 
$\Gamma$ for all $t\ge0$: write $G^0=\{g^0_t:t\ge0\}$. Of course, it is 
possible to write other derivatives on $X$ possessing some of the properties 
of options. In the Appendix we prove the following:

\begin{theorem}
\label{th price}
For each $f\in\Gamma$ there exists $\theta(f)\in\Theta$ such that 
$\pi\big(f(X)/(X\wedge1)\big)=q\big(\theta(f)\big)$ and
\begin{equation}
\label{qg}
q\big(\theta(f)\big)
	=
[q(j_0),\ldots,q(j_I)]
\left[
\begin{tabular}{cccc}
$(j_1-j_0)$&$0$&$\ldots$&$0$\\
$(j_2-j_0)$&$(j_2-j_1)$&$\ldots$&$0$\\
$\vdots$&$\vdots$&$\ddots$&$\vdots$\\
$d_0(\bar X)$&$d_1(\bar X)$&$\ldots$&$d_I(\bar X)$
\end{tabular}
\right]^{-1}
\left[
\begin{tabular}{c}
$f(j_1)$\\
$\vdots$\\
$f(j_I)$\\
$\hat f(\bar X)$
\end{tabular}
\right]
\end{equation}
where $d_i(\bar X)=(\bar X-j_i)$ and $\hat f(\bar X)=f(\bar X)$ if 
$\bar X<\infty$ or else $d_i(\bar X)=1$ and $\hat f(\bar X)=\lim_nf(n)/n$.
In particular, we can write
\begin{equation}
q\big(\theta(f)\big)
	=
\sum_{i=0}^Iw_iq(j_i)
	=
\sum_{i=1}^Ib_if(j_i)+b_{I+1}\hat f(\bar X)
\end{equation}
Let $D_i(h)=\frac{h(j_{i+1})-h(j_i)}{j_{i+1}-j_i}$ when $i=0,\ldots,I-1$ and 
$D_{-1}(h)=D_I(h)=0$. Then, if $\bar X=\infty$ we have 

\begin{subequations}
\label{wb}
\begin{equation}
w_I=\hat f(\bar X)-D_{I-1}(f),\quad b_{I+1}=q(j_I)
\end{equation}
\begin{equation}
w_i=D_i(f)-D_{i-1}(f)
\quad\text{and}\quad
b_{i+1}=D_{i+1}(q)-D_i(q)
\qquad
i=0,\ldots,I-1
\end{equation}
\end{subequations}
\end{theorem}

The next result is just a restatement of \cite[Theorem 7]{ThOpt}.

\begin{theorem}
\label{th representation}
Let $G=\{g_t:t\in\R_+\}$ be a collection of functions $g_t\in\Gamma$ such that 
\begin{equation}
\label{cdom}
\alpha g_s+(1-\alpha)g_u\ge g_t
\qquad\text{whenever}\quad
\alpha\in[0,1],\  
s,t,u\in\R_+
\quad\text{and}\quad 
\alpha s+(1-\alpha)u\le t
\end{equation}
Write $q^G(t)=\pi\big(g_t(X)/(X\wedge1)\big)$. There exists $\beta^G\ge0$
and $\nu^G\in ca\big(\Bor(\R_+)\big)$ such that
\begin{equation}
\label{option}
q^G(t)
	=
\beta^G+\int_t^\infty\nu^G(x>z)dz
\qquad\text{for all } t\ge0
\end{equation}
\end{theorem}

Observe that, by standard rules,
\begin{equation}
\int_t^\infty\nu^G(x>z)dz=\int(x-t)^+d\nu^G(x)
\end{equation}
Thus \eqref{option}, represents the price of the $G$ derivatives as the sum
of a bubble part and of their fundamental value. This view is substantiated by
the observation that necessarily
\begin{equation}
\beta^G
	=
\lim_{t\to\infty}q^G(t)
\end{equation}
so that, upon choosing $G=G^0$, the term $\beta^0$ represents the option price
as the strike approaches infinity and contributes to explaining the overpricing
of deeply out of the money CALL's often documented empirically in some form
of the smile effect.

As remarked in \cite{ThOpt}, Breeden and Litzenberger formula applies, giving:
\begin{equation}
\label{breeden}
\nu^G(x>t)=-\left.\dDer{q^G(k)}{k}\right\vert_{k=t}
\qquad\text{for all } 
t\ge0
\end{equation}
while in the model of Black and Scholes one has, as is well known,
\begin{equation}
\label{BS}
\nu^{BS}(x>k)=e^{-rT}\Phi(d_2)
\quad\text{with}\quad
d_2=\frac{\ln(S_0/k)+(r-\frac12\sigma^2)T}{\sigma\sqrt T}
\end{equation}
so that $\bnorm{\nu^{BS}}=\exp(-rT)$.  Observe that in our setting, we do not have an
\textit{a priori} restriction on the norm of $\nu^G$ because our market 
does not include any riskless bond.

\section{Estimating the CALL function}
\label{sec estimate}
% Add MSE or IMSE
%Draw true + error simulated values as in Y H Fig. 1

For what concerns the choice of $G$, it is natural to start considering the family 
$G^0$. By construction $q^0(k)$ represents the efficient price of the corresponding 
option and so  it coincides with $q(k)$ if and only if $k\in K^0(X)$. More generally 
one sees that the function $q^0$ is the highest among the positive, convex and 
decreasing curves passing through  the knots $\{(j,q(j)):j\in K^0(X)\}$. A\"it-Sahalia 
and Duarte \cite{ait sahalia duarte} obtain the same values in the first step of 
their approach. Since we have to make a choice for definiteness let's  assume
henceforth

\begin{assumption}
\label{ass infty}
$\bar X=\infty$.
\end{assumption}

Replace $g$ with $g^0_k$ in Theorem \ref{th price}. Then, according to \eqref{wb}, 
the cheapest way to superhedge the corresponding option is to invest
\begin{equation*}
w^0_i(k)=\frac{j_{i+1}-k}{j_{i+1}-j_i}
\quad\text{and}\quad
w^0_{i+1}(k)=1-w^0_i(k)
\end{equation*}
in the options with strike prices $j_i$ and $j_{i+1}$ respectively, where 
$j_{i+1}\ge k> j_i$, and nothing in all other options; or else, if $k>j_I$,
to buy one unit of the option with strike price $j_I$. The corresponding 
superhedging price would be
\begin{equation}
\label{q0}
q^0(k)=
\left\{
\begin{tabular}{lll}
$\frac{j_{i+1}-k}{j_{i+1}-j_i}q(j_i)+\frac{k-j_i}{j_{i+1}-j_i}q(j_{i+1})$&if&$j_i<k\le j_{i+1}$\\
$q(j_I)$&if&$j_I<k$
\end{tabular}
\right.
\end{equation}
Observe that the CALL function $k\to q^0(k)$ satisfies: (\textit{i}) 
$q^0(j_i)=q(j_i)$ for $i=0,\ldots,I$ and (\textit{ii}) it is a straight
line on each interval $[j_i,j_{i+1}]$ for $i=0,\ldots,I-1$ and on 
$[j_I,\infty)$. Then it coincides with the projection obtained by following 
Dykstra's method, see \cite[sec. 4.1]{dykstra}%
\footnote{
In fact, A\"it-Shalia and Duarte have to adapt slightly the method of
Dykstra since, assuming the lower bound for CALL options holds, the
set unto which they project is not a convex cone but just a convex set.
However, this additional restriction, for the reasons outlined above,
does not apply here. See \cite[p. 18 and Appendix A]{ait sahalia duarte}.
}.

From \eqref{q0} we deduce easily that
\begin{equation}
\label{nu0}
\nu^0(x>k)=
\left\{
\begin{tabular}{lll}
$\frac{q(j_i)-q(j_{i+1})}{j_{i+1}-j_i}$&if&$j_i<k\le j_{i+1}$\\
$0$&if&$j_I<k$
\end{tabular}
\right.
\end{equation}
In other words, the implied set function $\nu^0$ coincides -- not surprisingly --
with the derivative of the price function over the discrete set $K^0(X)$ upon
a change of sign.

Despite being a theoretically exact formula, the informational content of 
\eqref{nu0} is indeed quite poor empirically, as a consequence of the limited 
number of options which are priced efficiently by the market -- the size of 
$K^0(X)$. In fact $\nu^0(x>t)$ remains constant within two adjacent strike 
prices in $K^0(X)$ or, equivalently, the efficient CALL function $q^0(k)$ is 
piecewise linear. Adding the fictitious efficient prices $q^0(k)$ to the original 
data set, thus, does not improve our knowledge of $\nu$. 

Another way of putting it is saying that superhedging of options is an
intrinsically trivial exercise, a conclusion to which contribute two distinct
factors. First, the piecewise linear nature of the payoff function is such
that superhedging never requires more than two efficient options which
makes the corresponding price function locally linear, as in \eqref{q0}. On 
the other hand, the relative scarcity of available efficient prices makes the 
length of intervals on which the function is linear wider. If the size of the
set $K^0(X)$ is a constraint in this problem and essentially depends on
the market structure, a quick look at \eqref{qg} reveals that smoothness 
of $q^G$ depends on the smoothness of the underlying family $G$. Given 
that obtaining smooth estimates is the goal of any econometric exercises, 
our first step is then to construct a family $G^h$ of derivatives written on 
$X$, each having a payoff -- denoted by $g_k^h(X)$ -- which is (\textit{i}) 
conveniently close to the corresponding CALL option but (\textit{ii}) twice 
continuously differentiable and such that (\textit{iii}) $G^h$ satisfies 
\eqref{cdom}. These properties guarantee that the corresponding price, 
$q^h(k)$, will be a smooth estimate of the option prices and that it will
admit the representation \eqref{option} in terms of an implicit pricing
measure. The class $G^h$ of derivatives will depend on a control parameter, 
$h>0$, which acts in much the same way as the bandwidth parameter in 
kernel regression. We give now a more detailed description.

\subsection{The Spline Approach}

Let $k,h>0$ and $N\in\N$ be given. Divide the interval $[-h,h]$ into $N$ 
intervals of equal length, with endpoints $t^h_i=-h+2ih/N$ and $i=0,\ldots,N$. 
Consider then the following functional (with $D^2f$ denoting the second derivative 
of $f$ and $g^0_k$ the CALL payoff function with strike $k$ as above):
\begin{equation}
\label{I}
I_k(h;g;\lambda)
	\equiv
\sum_{i=0}^N\left[g^0_k\big(k+t_i^h\big)-g\big(k+t_i^h\big)\right]^2
	+
\lambda\int_{k-h}^{k+h}\left[D^2g(x)\right]^2dx
\end{equation}
and the program
\begin{equation}
\label{spline}
I_k(h;\lambda)
	\equiv
\min_{g\in\Gamma\cap\ \mathscr C^2}I_k(h;g;\lambda)
\end{equation}
It is well known, see \cite[Theorem 5.2]{eubank}, that a solution to this problem 
is given by a $\mathscr C^2$ cubic spline which is linear outside of $[k-h,k+h]$. 
Based on the fact that the second derivative of a cubic spline is locally linear, the 
infinite dimensional problem \eqref{spline} conveniently reduces to a $2N$%
-dimensional one. Turlach \cite{turlach} developed a methodology to compute 
its solution under several shape restrictions such as (\textit i) $g,Dg,D^2g\ge0$, 
i.e. positivity, monotonicity and convexity. To these constraints we add the 
following: (\textit{ii}) $g(x)=Dg(x)=0$ for $x\le k-h$, (\textit{iii}) 
$g(k+h)=g^0_k(k+h)$ and (\textit{iv}) $Dg(x)=1$ for $x\ge k+h$. Denote
\begin{equation}
\label{constraints}
\chi(k;h)
	=
\left\{
g\in\mathscr C^2(\R_+):g\text{ cubic spline meeting the constraints }(i)-(iv)
\right\}
\end{equation}

The notation adopted is consistent with the choice of treating $N$ as a fixed
parameter and of focusing exclusively on properties which depend on $h$.
Although we experimented several values, in what follows we will set $N=10$.
More importantly, we make the choice of $\lambda$ endogenous by letting
$\lambda_h=(0.1h)^3$ -- so that $\lambda_{10}=1$. The existence of a 
constrained solution to \eqref{spline} and the properties of such solution are 
proved in the next:

\begin{lemma}
\label{lemma turlach}
The problem 
\begin{equation}
I_k(h)
	\equiv
\min\big\{I_k(h;g;\lambda_h):g\in\chi(k;h)\big\}
\end{equation}
admits one and only one solution, $g_k^h\in\chi(k;h)$, 
and this satisfies: 
(i) 
$g^h_{k+m}(x)=g^h_k(x-m)$ for all $k,m,x\ge0$,
(ii) 
$g_k^0\le g_k^{h'}\le g_k^h$ whenever $h'\le h$
and 
(iii) 
$\lim_{h\to0}\sup_x\big(g^h_k-g^0_k\big)(x)=0$. 
\end{lemma}

Thus if we set
\begin{equation}
G^h=\big\{g_k^h:k\ge0\big\}\subset\Gamma
\end{equation}
the family $G^h$ satisfies \eqref{cdom}. 
Actually, in the proof of Lemma \ref{lemma turlach} we obtain that
for fixed $k_0,h_0>0$
\begin{equation}
\label{spline projection}
g_k^h(x)
	=
g_{k_0}^{h_0}\left(k_0+(x-k)\frac{h_0}{h}\right)\frac{h}{h_0}
\qquad\text{for every}\quad
k,\ h>0
\end{equation}
so that property (\textit{iii}) of Lemma \ref{lemma turlach} follows easily from

\begin{align}
\label{qh to q0}
0\le q^h(k)-q^0(k)
	\le
\pi\left(\frac{g^h_k(X)-g^0_k(X)}{X\wedge1}\right)
	\le
\frac{g_k^h(k)}{(k-h)\wedge1}q(0)
	\le
\frac{g_k^{h_0}(k)q(0)}{(k-h_0)\wedge1}\ \frac{h}{h_0}
\end{align}
Denote by $\theta_k^h\in\Theta$ the portfolio $\theta(g_k^h)$ involved in
super replicating $g_k^h$ and by $\hat g_k^h$ its payoff.

\begin{center}
\begin{figure}
\tiny{Panel A: Option Payoff $g_k^h(X)$}\\\vspace{-0.5cm}
\includegraphics[width=0.8\textwidth]{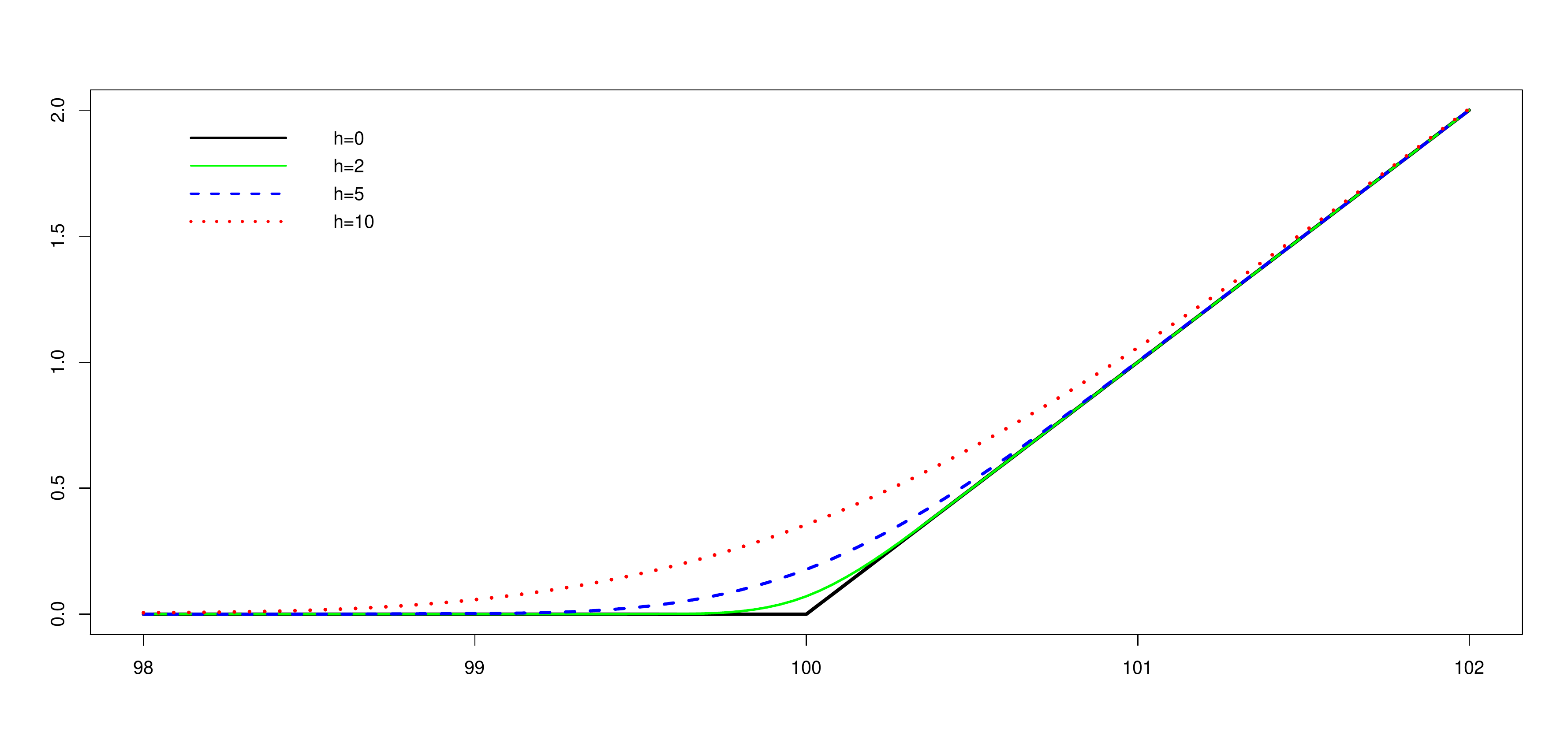}\\

\tiny{Panel B: Option Payoff Derivative $Dg_k^h(X)$}\\\vspace{-0.5cm}
\includegraphics[width=0.8\textwidth]{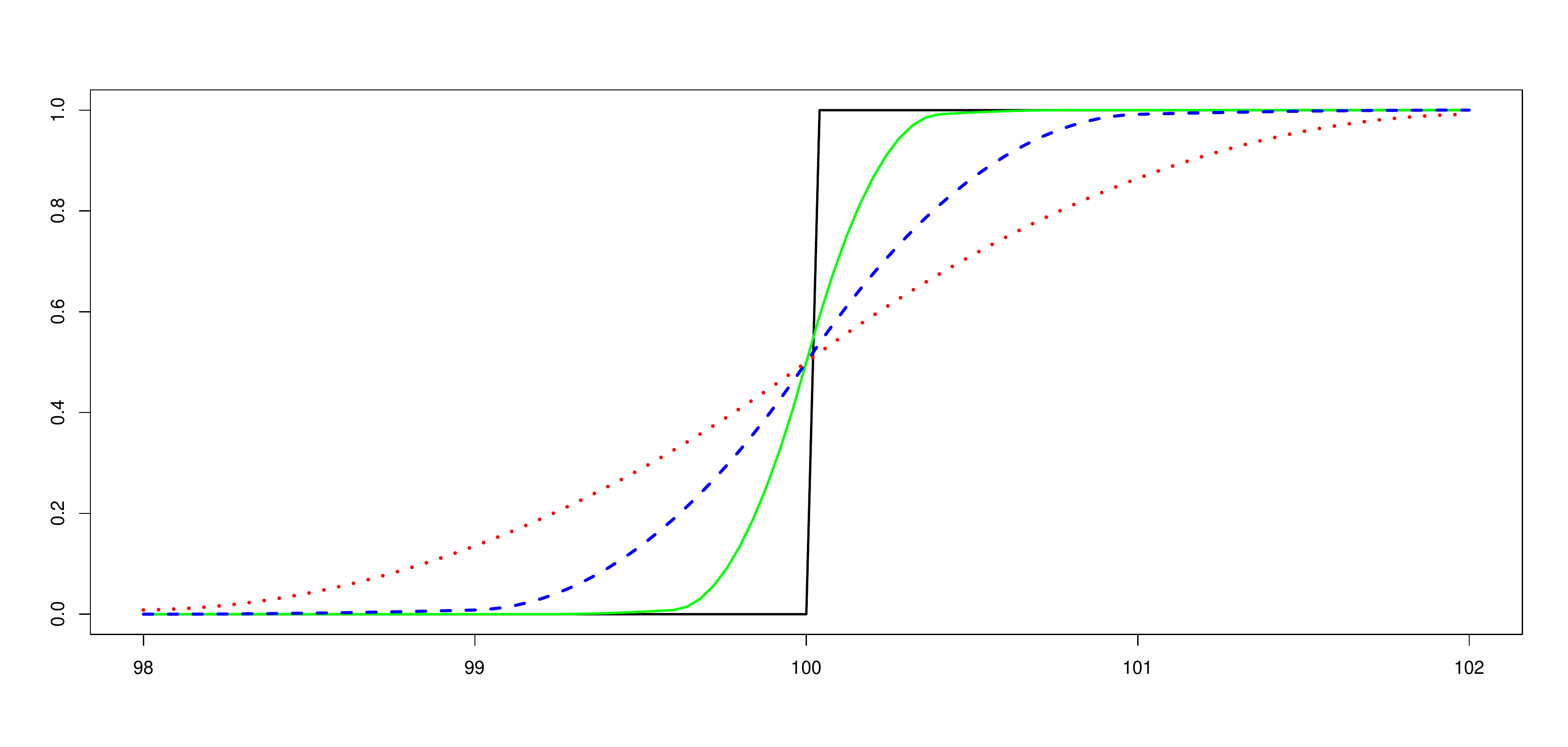}\\

\tiny{Panel C: Smooth Superhedging}\\\vspace{-0.5cm}
\includegraphics[width=0.8\textwidth]{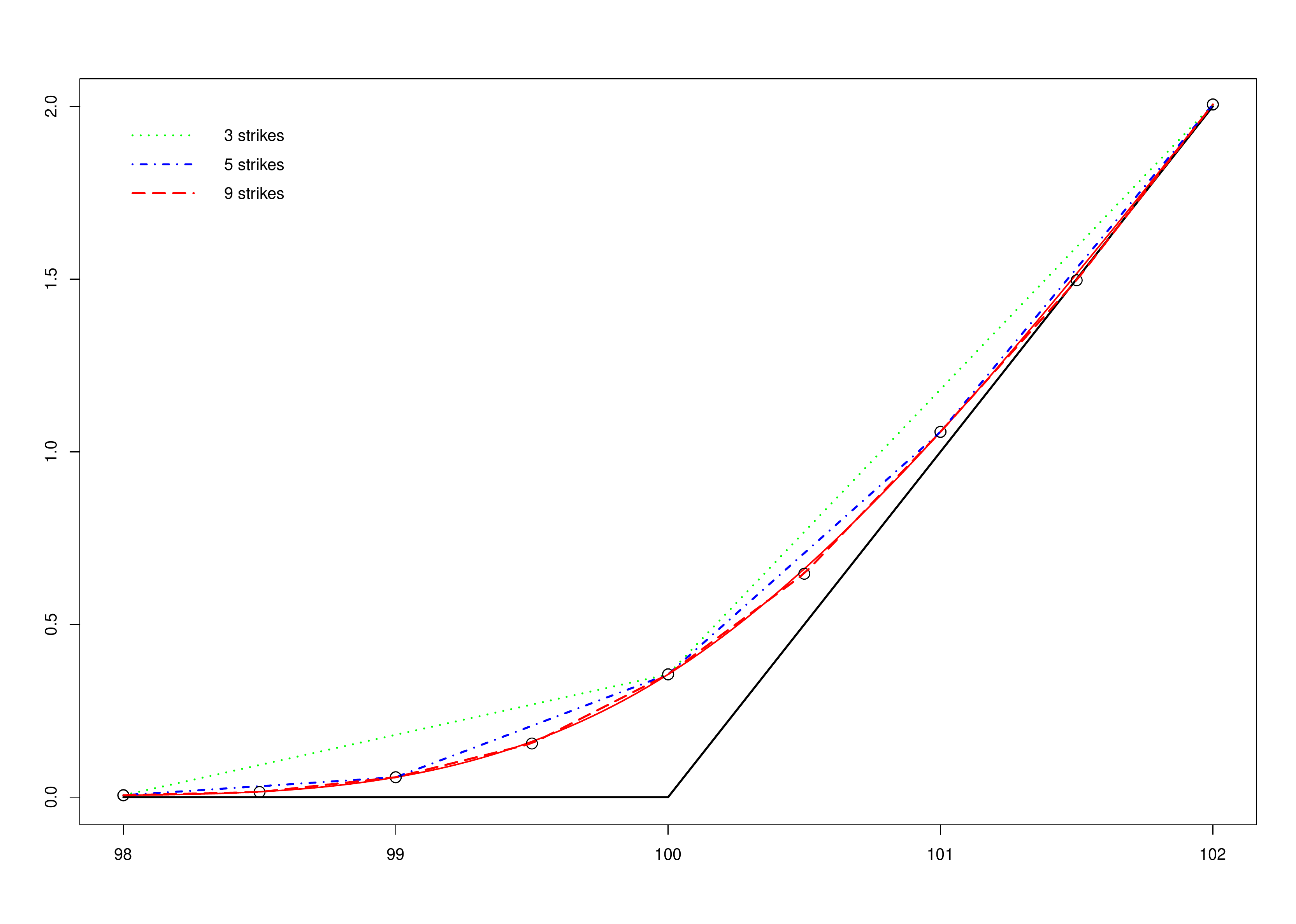}

\caption{\label{fig spline}
Smoothing Option Payoff by $h$.\\
{\tiny
Plot of $g_k^h(x)$ (Panel $A$) and $Dg_k^h(x)$ (Panel $B$) for $k=100$ 
and $h=2,5,10$. In Panel C we draw $g_k^{h=10}(x)$ (red dotted line)
together with $\hat g_k^{h=10}(X)$ (piecewise linear curves), the payoff of 
the portfolio superhedging it assuming to have 3, 5 or 9 equally spaced strike 
prices.
}
}
\end{figure}
\end{center}

The content of Lemma \ref{lemma turlach} is clearly illustrated in Figure 
\ref{fig spline}, Panels A and B, where the function $g_k^h$ and its derivative 
are plotted for different values of $h$.

In Panel C the target payoff $g_k^h$ is plotted together with the payoff 
$\hat g_k^h$ of the portfolio that super replicates it in order to understand 
the role of efficiency. By construction, superhedging $g_k^h$ or $\hat g_k^h$ 
are entirely equivalent exercises. Thus, the degree of smoothness involved 
depends not only on the shape of the notional payoff $g_k^h$ but also on 
that of its market counterpart, $\hat g_k^h$, and particularly on the number 
of contracts involved in superhedging. This remark highlights the role of the 
bandwidth $h$. A small value of this parameter makes the width of the 
smoothing interval $[k-h,k+h]$ narrow and diminishes, as a consequence, 
the distance between $g_k^0$ and $g_k^h$, as shown in Lemma 
\ref{lemma turlach}. At the same time the number of efficiently priced 
options with strike price included in such interval -- i.e. the size of the set 
$[k-h,k+h]\cap K^0(X)$ -- is reduced, making the super 
hedging payoff $\hat g_k^h$ less smooth. In fact, in Panel C -- with $6$ 
equally spaced efficient strike prices ranging from $95$ to $105$ -- the payoff 
$g_k^h$ is represented by a continuous, smooth line while $\hat g_k^h$ 
is the piecewise linear curve of the same color dominating it. In the case 
$h=2$, the super hedging portfolio only contains $2$ options while for the 
case $h=10$ it contains all $6$. It is clear from the picture that $\hat g_k^h$ 
becomes less smooth as it approaches $g_k^0$, suggesting the need to 
balance these two countervailing effects.

To understand this point better, let $M$ be the mesh of $K(X)\setminus\{0\}$, 
i.e.
\begin{equation}
\label{mesh}
M
	=
\sup\big\{\abs{k-k'}:k,k'\in K(X)\setminus\{0\},\ k\ne k'\big\}
\end{equation}
and $M^0$ the mesh of $K^0(X)\setminus\{0\}$. Let also $w_k^h(j)$ denote the quantity 
invested in option $\theta(j)$, with $j\in K^0(X)$, in order to super hedge 
$g_k^h$, obtained from \eqref{wb}. From the fact that $g_k^h$ and $g_k^0$
coincide outside of the interval $[k-h,k+h]$ by \eqref{constraints} and from 
\eqref{wb} we deduce the implications
\begin{subequations}
\label{w>0}
\begin{equation}
\label{w>0,i>0}
w_k^h(j_i)>0
\quad\Rightarrow\quad
j_{i+1}+h>k>j_{i-1}-h
\quad\Rightarrow\quad
\abs{k-j_i}< h+M^0
\qquad
i=1,\ldots,I
\end{equation}
\begin{equation}
\label{w>0,i=0}
w_k^h(0)>0
\quad\Rightarrow\quad
j_1>k-h
\end{equation}
\end{subequations}
This suggests to set $h=\delta M$ so to make the properties of the estimator 
sample dependent. In the sequel we will replace the superscript $h$ with 
$\delta$. We observe that, as a consequence of \eqref{w>0}, the number of 
efficient options employed in order to superhedge $g^\delta_k$ is at most 
$3+2\delta M/M^0$ so that it is desirable to fix $\delta>M^0/M\ge1$ in order 
to have at least $5$ options available. In the following sections we will experiment 
values of $\delta$ ranging from $1$ to $10$.

\subsection{Properties of the Estimator}

A different and more classical question is whether the proposed estimator 
converges to the \textit{true} price function in the presence of disturbances. 
The classical formulation of this problem is
\begin{equation}
\label{model}
q(k)=F(k)+\varepsilon_k
\qquad k\in K(X)
\end{equation}
where $F$ is the model, and is thus assumed to be a positive, decreasing, convex,
$\mathscr C^2$ function, whereas the errors $\varepsilon_i$ are identically and
independently distributed with zero expectation.

Theorem \ref{th price} gives an explicit functional form to our estimator:
\begin{equation}
\label{estimator}
q^\delta(k)
	=
\sum_{j\in K^0(X)}w_k^\delta(j)q(j)
\qquad
t>0
\end{equation}
Observe that, by \eqref{w} and Assumption \ref{ass infty}, 
\begin{equation}
w_k^\delta(j)\ge0
\quad\text{for all}\quad j\in K^0(X)
\quad\text{and}\quad
\sum_{j\in K^0(X)}w_k^\delta(j)=1
\end{equation}
At first sight, then, our estimator appears as an exemplification of the local 
averaging approach which includes, as additional special cases, kernels and 
regressograms. For this class of estimators a well established theory 
demonstrates (see e.g. \cite[p. 677]{yatchew}) that the \textit{MSE} 
converges to $0$ as the sample size diverges. However, in local average
estimators weights are assumed to be uncorrelated with disturbances and
this is of crucial importance in proving convergence. In our setting, instead,
this property can in no way be assumed to hold. In fact the set $K^0(X)$
of efficient option prices, hitherto treated as given, will in general strongly
depend on prices and therefore on disturbances. Not only but the inclusion
of a given price $k$ in $K^0(X)$ will depend not only on the corresponding noise 
but also on all other disturbances since a high value for $\varepsilon_{k'}$ 
will make it more likely for $q(k)$ to be efficient. The problem discussed 
provides a clear exemplification of the selection bias studied by Heckman 
\cite{heckman} and it is to some extent surprising that its pervasive role
in empirical option pricing has not been fully recognized. From the classical
work of Heckman we learn, in fact, that selection effects may result in
biased and inconsistent estimates unless correcting for a term which proxies 
the expected value of disturbances conditional on selection. The work of
Heckman has been extended to the non parametric setting in a recent paper
by Das et al. \cite{das newey vella}. Our aim is that of giving sufficient conditions 
for the $MSE$ to converge to $0$ when taking the selection effect fully into account.

The first step is to treat $K^0(X)$ as a random variable. Observe to this end
that the inclusion $k\in K^0(X)$ holds if and only if the $\F$ measurable event 
$A_k=\{q(k)\le q^0(k)\}$ occurs. In order to take randomness into full account
we replace the symbol $K^0(X)$ with $Z(\omega)\subset K(X)$. Then, if 
$z\subset K(X)$ we have that $\{Z=z\}=\bigcap_{k\in z}A_k\in\F$. We
observe that although the value of $w_k^\delta(j)$ depends crucially on $Z$, 
it is in fact deterministic once conditional on the event $\{Z=z\}$. It is thus 
natural to follow a two step procedure (similarly to Heckman and to Das et al.) 
by first conditioning all variables on the selection mechanism and then focusing 
on the unconditional properties. Denote by $P_z(\cdot)$ the conditional
 expectation given the event $\{Z=z\}$, for $z\subset K(X)$ and
$P_Z(\cdot)=\sum_{z\subset K(X)}P_z(\cdot)\sset{Z=z}$. When needed
we explicit $z$ as $\{0=j_0^z<j_1^<\ldots<j_{I^z}^z\}$ and write
\begin{equation*}
\tau_i=\sum_{z\subset K(X)}j_i^z\sset{Z=z}
\end{equation*} 
Denote by $M_z$ and $M_Z$ the mesh of the sets $z\setminus\{0\}$ and 
$Z\setminus\{0\}$ respectively. The variable $P_Z(\varepsilon)$ will be our 
correction term.

\begin{theorem}
\label{th convergence}
Let the function $F:\R_+\to\R_+$ in \eqref{model} be decreasing and convex
and assume that $F\in\mathscr C^3$ with $\sup_{x\in\R_+}\abs{D^3F(x)}<\alpha$.
Assume moreover that
\begin{subequations}
\label{ass}
\begin{equation}
\label{ass a}
E(\varepsilon_\tau\varepsilon_\sigma)
	=
0,
\quad
P_Z(\varepsilon_\tau)
	=
P_z(\varepsilon_\sigma)
\quad\text{and}\quad
P_Z(\varepsilon_\sigma\varepsilon_\tau)
	=
P_z(\varepsilon_{\sigma'}\varepsilon_{\tau'})
\qquad\text{for all} \quad
\sigma,\tau,\sigma',\tau'\in Z
\end{equation}
\begin{equation}
\label{ass b}
E\Big(\sup_{\tau\in Z}P_Z(\abs{\varepsilon_\tau})\Big)<\infty
\quad\text{and}\quad
E\Big(\sup_{\sigma,\tau\in Z}P_Z(\abs{\varepsilon_\sigma\varepsilon_\tau})\Big)
	<
\infty
\end{equation}
\begin{equation}
\label{ass c}
\lim_{M\to0}E\big(M_Z^6\big)=0
\quad\text{and}\quad
\lim_{M\to0}P(\tau_1>0)=0
\end{equation}•
\end{subequations}
Then, for any compact interval 
$I\subset\R_+\setminus\{0\}$
\begin{equation}
\label{convergence}
\lim_{M\to0}\sup_{t\in I}E\Big(\big(q^\delta(t)-F(t)\big)^2\Big)=0
\end{equation}
\end{theorem}

Condition \eqref{ass a} is conceptually akin to the assumption made by Das et 
al. that the conditional expectation of the disturbance given selection depends 
only on the propensity score, see \cite[Assumptions 2.1, 2.2 and 2.3]{das newey vella}. 
Condition \eqref{ass b} is easily satisfied if, e.g., disturbances are uniformly 
distributed. The the most delicate property is definitely \eqref{ass c} as it
requires a strong form of convergence to $0$ of the mesh of $K^0(X)$. 
The proof of Theorem \ref{th convergence} strongly relies on some properties 
of the smoothing spline $g_k^\delta$ proved in Lemma \ref{lemma turlach}, a 
fact emphasizing the role of splines.

\subsection{The density approach}

As mentioned in the introduction, splines are not the only possible choice for
smoothing option payoff. We present in this paragraph an alternative based on
some given probability density function, $\phi$, satisfying the properties 
\begin{align}
\label{int}
\int u\phi(u)du=0
\quad\text{and}\quad
\int\abs u\phi(u)du<\infty
\end{align}
Denote by $\Phi$ the distribution function associated with $\phi$ and consider 
the following function:
\begin{equation}
\label{density}
g_k^h(x;\phi)
	=
\Phi\Big(\frac{x-k}{h}\Big)(x-k)-h\int_{-\infty}^{(x-k)/h}u\phi(u)du
\end{equation}

\begin{lemma}
\label{lemma density}
Let $\phi$ be a probability density function satisfying \eqref{int}. Then the
function $g_k^h(\cdot;\phi)$ defined in \eqref{density}	satisfies properties
(i), (ii) and (iii) of Lemma \ref{lemma turlach}.
\end{lemma}

The proof of the Lemma is rather clear given the explicit form of the derivatives
of $g_k^h(x;\phi)$, i.e.
\begin{equation}
\label{derivatives}
Dg_k^h(x;\phi)
	=
\Phi\Big(\frac{x-k}{h}\Big)
\quad\text{and}\quad
D^2g_k^h(x;\phi)
	=
\phi\Big(\frac{x-k}{h}\Big)\frac1h
\end{equation}
in which it is implicit the inequality
\begin{align*}
0
	\le
g_k^h(x;\phi)-g_k^0(x)
	\le
g_k^h(k;\phi)-g_k^0(k)
	\le
h\int \abs u\phi(u)du
\end{align*}

A special case of \eqref{density} is given by the standard normal distribution:
\begin{equation}
\label{normal}
g_k^h(x;\mathscr N)
	=
\Phi_\mathscr N\Big(\frac{x-k}{h}\Big)(x-k)+h\phi_\mathscr N\Big(\frac{x-k}{h}\Big)
\end{equation}
that will be briefly considered in the applications that follow, as a term of comparison.

A noteworthy implication of \eqref{density}, emerging clearly from \eqref{derivatives}
and \eqref{qg}, is the possibility to extract from the non parametric prices
$q^h(k;\phi)$ an implicit risk neutral density in closed form, that is
\begin{equation}
\label{dphi}
d^h(x;\phi)
	=
\sum_{i=1}^Ib_iD^2g_{j_i}^h(x;\phi)
	=
\sum_{i=1}^Ib_i\phi\Big(\frac{x-j_i}{h}\Big)\frac1h
\end{equation}
where the parameters $b_i$ are the positive coefficients in \eqref{wb}%
\footnote{
In fact, when $f=g_k^h(x;\phi)$ and $\bar X=\infty$, then $\hat f(\bar X)=1$.
}.
The implicit risk neutral density belongs thus to a preassigned family of 
density mixtures without actually having to make this assumption but 
rather as the consequence of the exact market pricing of a specific class 
of derivatives, $g^h_k(\phi)$. Not only, but the mixing parameters
are fixed by the market so that in order to estimate the implied risk
neutral density one only has to specify the bandwidth $h$.

In the option pricing literature a lot of interest was raised by models
which assume  that the risk neutral density is a mixture of log normals. 
This modeling choice has been adopted, among others, by Ritchey 
\cite{ritchey}, Melick and Thomas \cite{melick thomas}, S\"oderlind 
and Svensson \cite{soderlind svensson} and S\"oderlind \cite{soderlind}. 
In \cite{soderlind}, S\"oderlind lists among the advantages of this 
approach the closed form for option prices and the flexibility of this family 
of densities which easily accommodates for skewness and kurtosis of 
returns while avoiding the difficulties inherent in the non parametric approach. 
Although this is clearly not the focus of our work, we mention however the 
relative ease of \eqref{dphi} in which the weights $b_i$ are in fact fixed by 
the market and do not have to be estimated, which is typically a delicate 
step involving the EM algorithm. In the next section we confront the implicit 
density \eqref{dphi} with the true one when it is assumed that the returns 
follow a mixture of log normals to provide at least some evidence that our 
approach is a viable solution for those cases in which the assumption of a 
mixture of log normals is justified.

\section{Empirical Applications: Simulation Analysis}
\label{sec simulation}
In this section we are going to test our methodology using as our benchmark
the classical model of Black and Scholes, augmented for smile effects, using
exactly the same values as in A\"it-Sahalia and Duarte \cite{ait sahalia duarte} 
for ease of comparison. The current underlying price is set at $1,365$, maturity 
to 3 months, the interest rate at $4\%$ and we assume no dividends. We consider 
25 strike prices equally spaced between $1,000$ and $1,700$, so that $M=28$. 
Volatility will be a linear, decreasing function of the strike, ranging from $40\%$ 
to $20\%$. With these values, the option prices range from $378$ to $0.9$. 

\subsection{Deterministic Analysis}

\begin{center}
\begin{figure}
\tiny{Panel A: Option Price $q^{\delta M}(k)$}\\\vspace{-0.5cm}
\includegraphics[width=\textwidth]{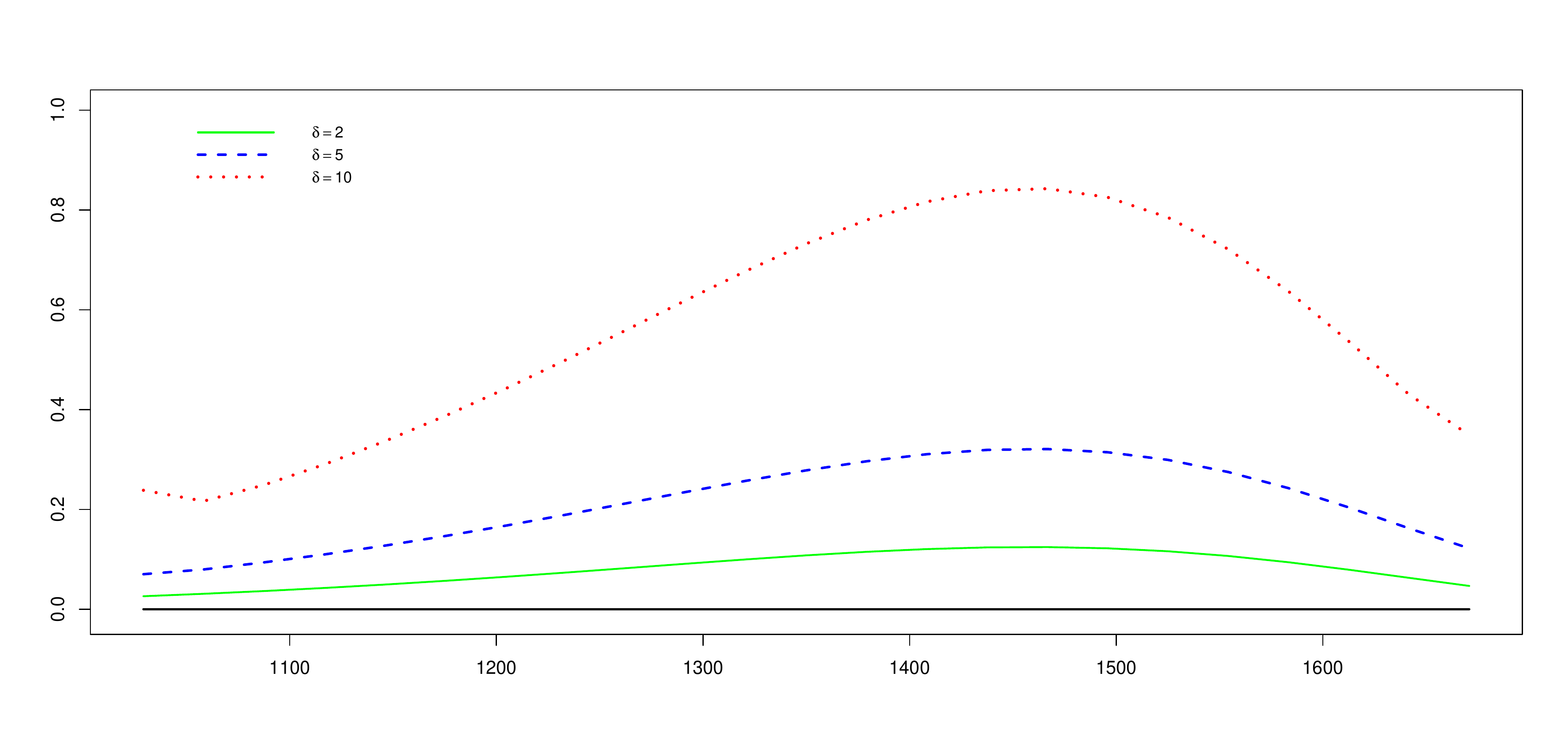}\\

\tiny{Panel B: Implied Volatility}\\\vspace{-0.5cm}
\includegraphics[width=\textwidth]{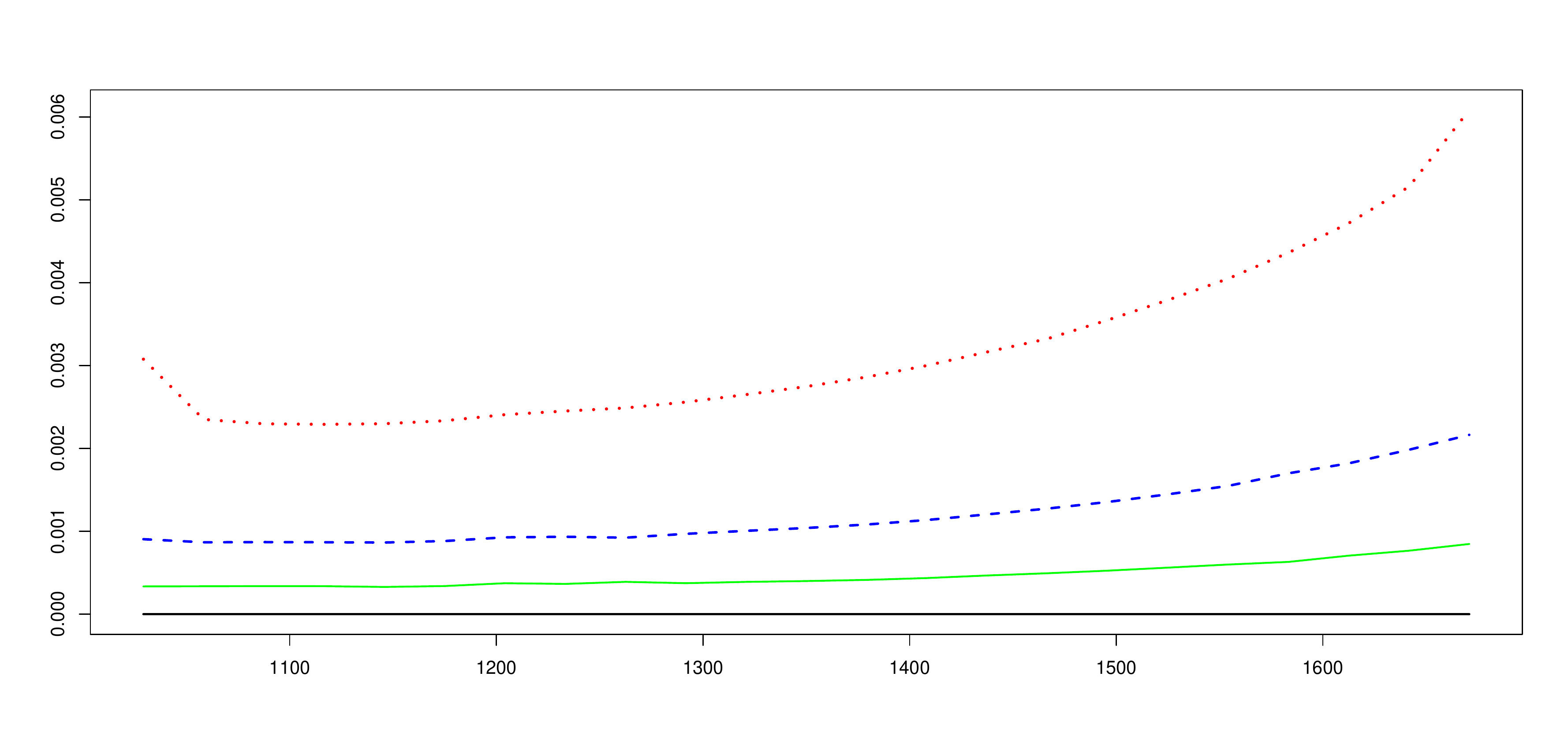}\\

\tiny{Panel C: Probability $\nu^{\delta M}(X>k)$}\\\vspace{-0.5cm}
\includegraphics[width=\textwidth]{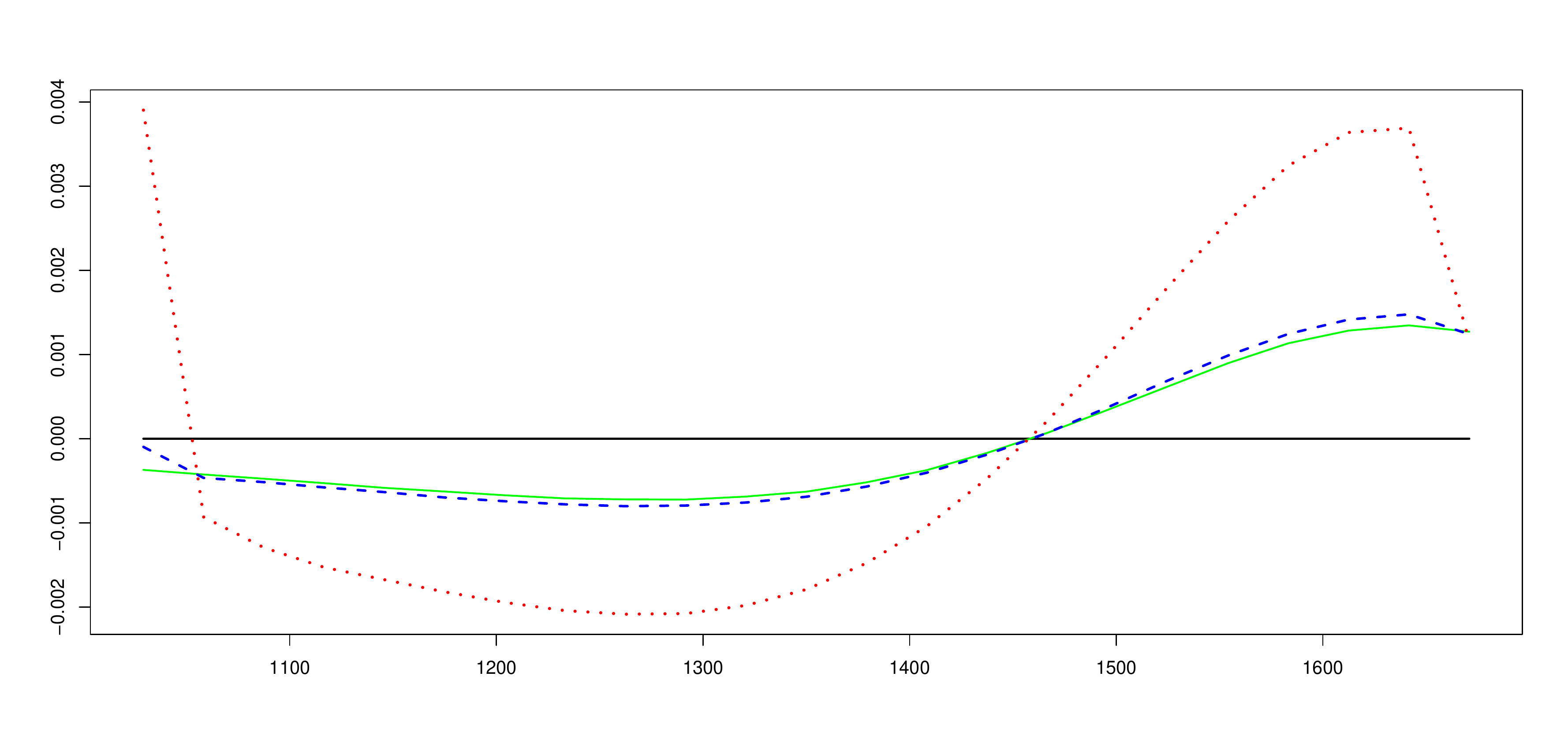}

\caption{\label{fig BS smile}
Deviations from Black and Scholes with a Smile.\\
{\tiny
$\sigma=0.4-0.2(k-1000)/700$, 
$\nu^{B\& S}=-\PD{q^{B\&S}}{K}-\PD{q^{B\&S}}{\sigma}\Der{\sigma}{k}$
and $N=10$.
}
}
\end{figure}
\end{center}

As a first step in our analysis we investigate how close  are the quantities 
$q^\delta(k)$, $\sigma^\delta(k)$ and $\nu^\delta(X>k)$ to the true values 
computed from the model, where $\sigma^\delta(k)$ is the the level of volatility 
implicit in $q^\delta(k)$ according to the Black and Scholes formula. We 
experiment three possible values of the smoothing parameter, namely 
$\delta=2,5,10$. The results, in terms of deviations of estimated values from 
true ones, are plotted in Figure \ref{fig BS smile}. Panel $A$ represents the 
price gap, Panel $B$ the volatility gap and Panel $C$ the gap in terms of 
cumulative probability. For all choices of $\delta$, the distance between true 
and fictitious values appear to be rather small%
\footnote{
Of course, the Black and Scholes implicit probability, $\nu^{B\&S}$, changes 
accordingly to incorporate the effect of the strike on volatility. 
}. 
Deviations of the smooth price from the true one are indeed quite limited
but contain a smile effect and increase for options out of the money.
This phenomenon, quite limited, soon disappears when we move from 
$\delta=10$ to $\delta=5$. Deviations from the benchmark are more 
interesting when considering the implicit probability as the smooth 
functions lie below $0$ up to $1,460$ for all choices of $\delta$.

\subsection{Smoothness vs. Variance}
The picture changes significantly as we introduce noise into the original 
model. To avoid arbitrariness we still follow A\"it-Sahalia and Duarte 
\cite{ait sahalia duarte}. In particular we believe that noise has an
implicit microstructural component that may be well captured by liquidity
and the bid/ask spread. We proxy illiquidity of options via the factor 
$1+5\abs{\exp(r\tau)k/S_0-1}$ which is 1 for options exactly at the money 
and gets as high as $2.3$ for deeply in and out of the money options. 
We fix the basis spread to $5\%$ of the option price with a floor at $50$ 
cents and a cap at $3$ dollars. Noise is then modeled as a random
variable which is uniformly distributed on an interval centered at the
origin and with radius equal to the product of illiquidity and half of the
spread.

\begin{center}
\begin{figure}
\includegraphics[width=1\textwidth]{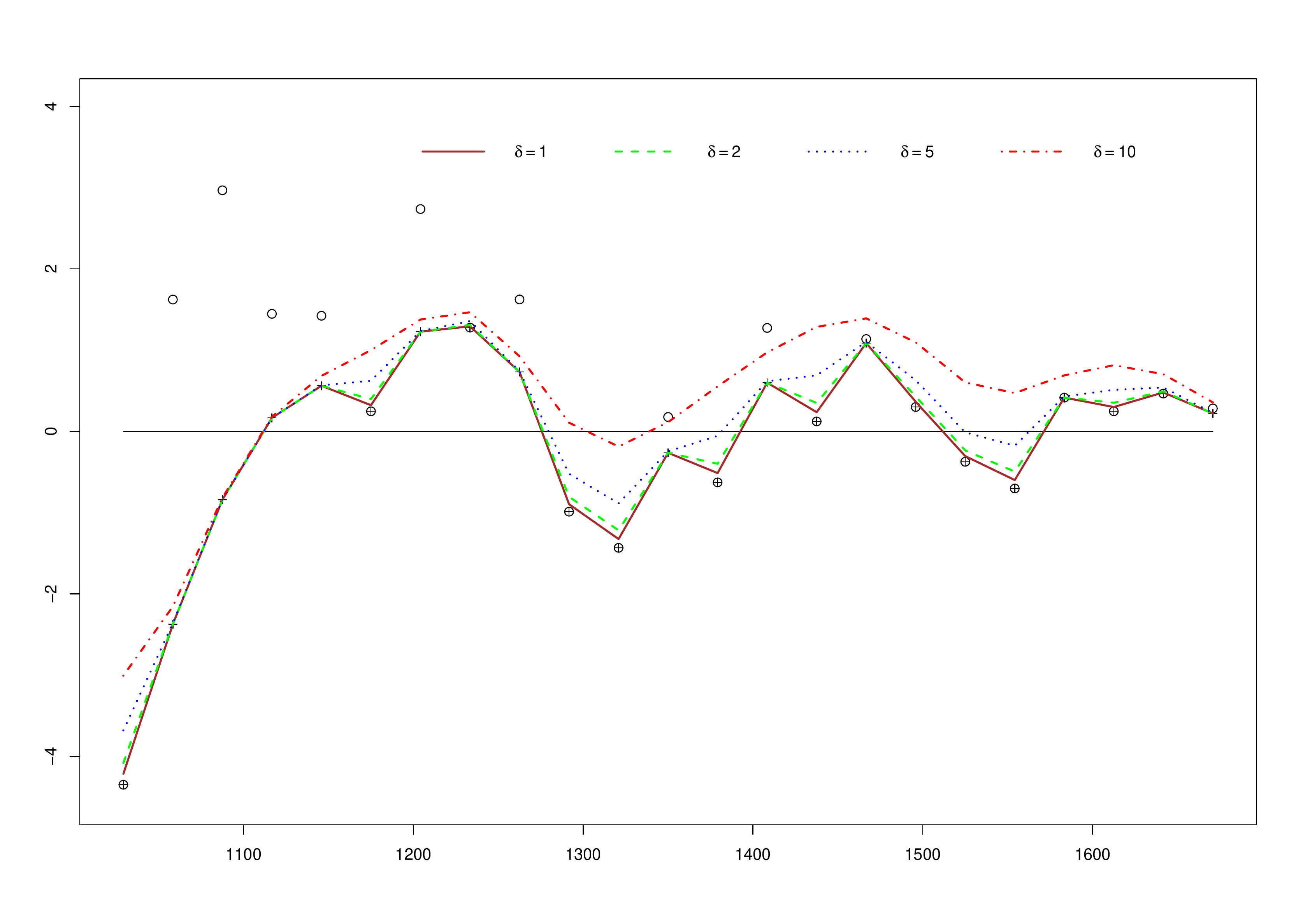}
\caption{\label{fig SmoothVar}
Smoothness vs. Variance.\\
{\tiny
	Estimating the model $q(k)+\varepsilon_k$ with $N=10$ and
$\delta=1,2,5,10$. Values are expressed in terms of their distance from
the correct B\&S price. Actual option prices are plotted as circles while
the corresponding efficient prices are plotted as crosses. The two
symbols are overwritten whenever the corresponding strike belongs
to $K^0(X)$.
}
}
\end{figure}
\end{center}

In Figure \ref{fig SmoothVar} we draw the distance between the correct
Black and Scholes price and its estimate $q^\delta(k)$ when prices are
affected by errors, in the way described above.

The picture shows that the higher the value of $\delta$ the more smooth 
is the resulting estimate while, at the same time, goodness of fit decreases. 
This is a classical finding in non parametric statistics but presents here some 
new feature. We observe first that non efficient option prices, corresponding 
to plain circles, have actually no impact on estimates. The corresponding 
superhedging prices are plotted as crosses (so that the superposition of a 
cross to a circle signifies that the corresponding strike is an element of 
$K^0(X)$). $8$ prices out of $24$ are non efficient and, as can be seen, 
most of them are concentrated on the left-hand side corresponding to deeply 
ITM options. This is due to the combined effect of illiquidity and of the spread 
which, in our set-up as in the real world, magnifies noise for this segment of 
the options market. We remark that noise has an asymmetric effect. A positive
shock, particularly in the presence of a strong microstructural multiplier, will
make the corresponding price inefficient and thus has no impact on our 
estimates. If the shock is negative or small in magnitude then it is less
likely to produce price inefficiency and will be included in the estimated values.
Our method is subject to a limited underestimation error for those prices which
are affected by errors in a more relevant way,as is the case here for options
deeply ITM. Second, we notice that the higher is $\delta$ the more upward 
shifted will be the fitting curve. This follows from the payoff $g_k^\delta$
being larger. Including a constant term to the estimated CALL function would 
indeed reduce this problem as long as the constant make take on either
sign. This addition would correspond however, in its financial counterpart, to the
possibility of taking long or short positions in the riskless asset, a modeling
choice which, although completely standard in the literature, blatantly contrasts
with our starting assumptions.

\subsection{Monte Carlo Simulation}
In order to construct interval estimates and evaluate the statistical aspects of
our approach we run $5,000$ Monte Carlo simulations of the error terms.
Our model takes the form
\begin{equation}
q(k)+\varepsilon_k^s,
\qquad
k\in K(X),\ s=1,2,\ldots,5000
\end{equation}
where the error terms are distributed as described above.
\\
\begin{center}
  \begin{tabular}{c|c|c|c|c|c|c}
\hline\hline
	&dITM	&ITM	&ATM	&OTM	&dOTM	&Total\\\hline
Nr	&6		&5		&5		&4		&5		&25\\
mean&52.12	&50.8	&30.5	&18.0	&15.32	&34.74\\
min	&0.0		&0.0		&0.0		&0.0		&0.0		&12.0\\
max	&83.33	&80.0	&80.0	&75.0	&60.0	&56.0%\hline\hline
  \end{tabular}
\nobreak\\\vspace{0.6cm}\nobreak
Table 1: Percentage of inefficient prices for each market segment.
\end{center}
The introduction of noise produces a number of arbitrage violations, as 
documented in the preceding Table 1. These violations amount on average 
to $34.74\%$ of the sample but range up to $56\%$ and are above 
$40\%$ in the $20\%$ of cases.

\begin{center}
\begin{figure}
\tiny{Panel A: Option Price}\\\vspace{-0.5cm}
\includegraphics[width=\textwidth]{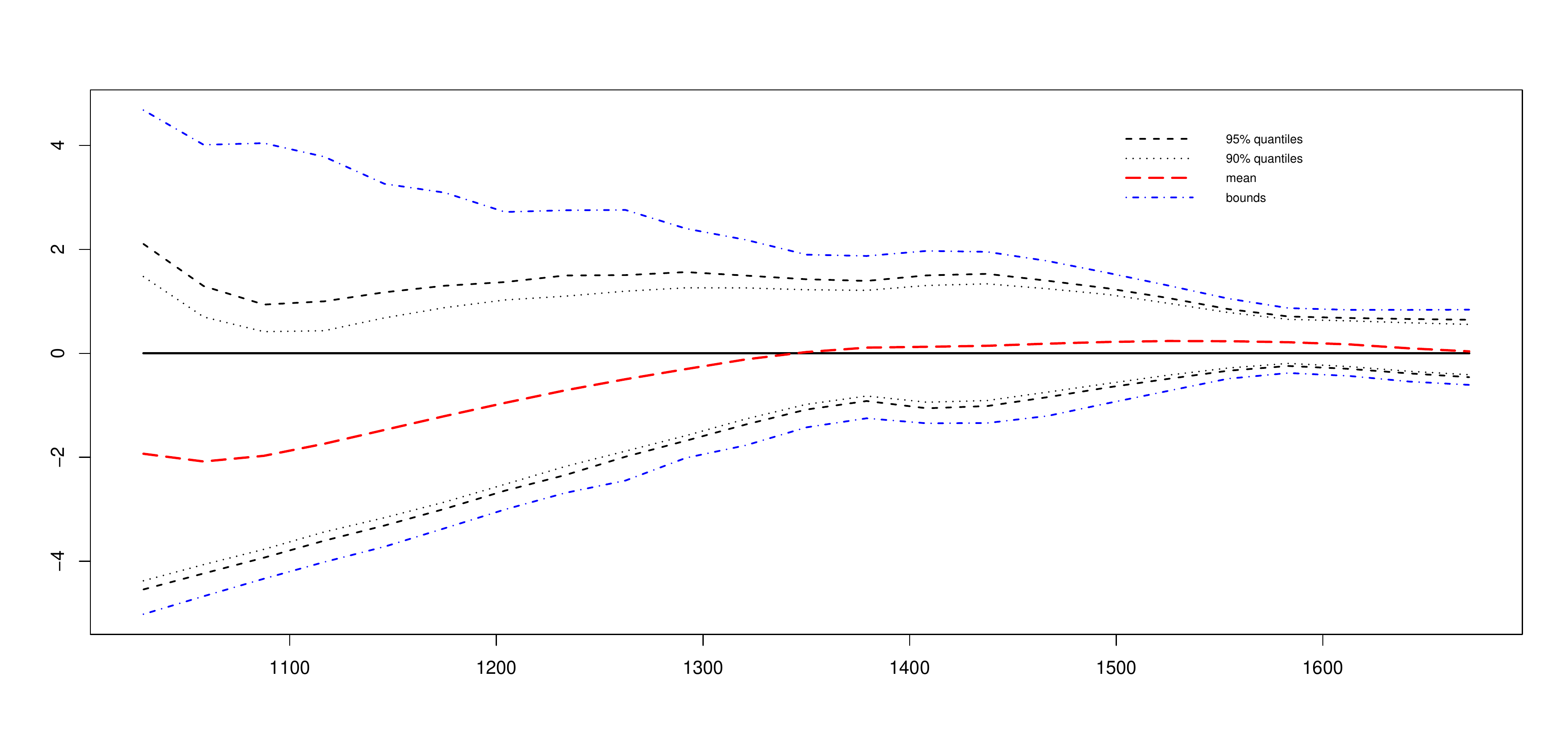}\\

\tiny{Panel B: Volatility}\\\vspace{-0.5cm}
\includegraphics[width=\textwidth]{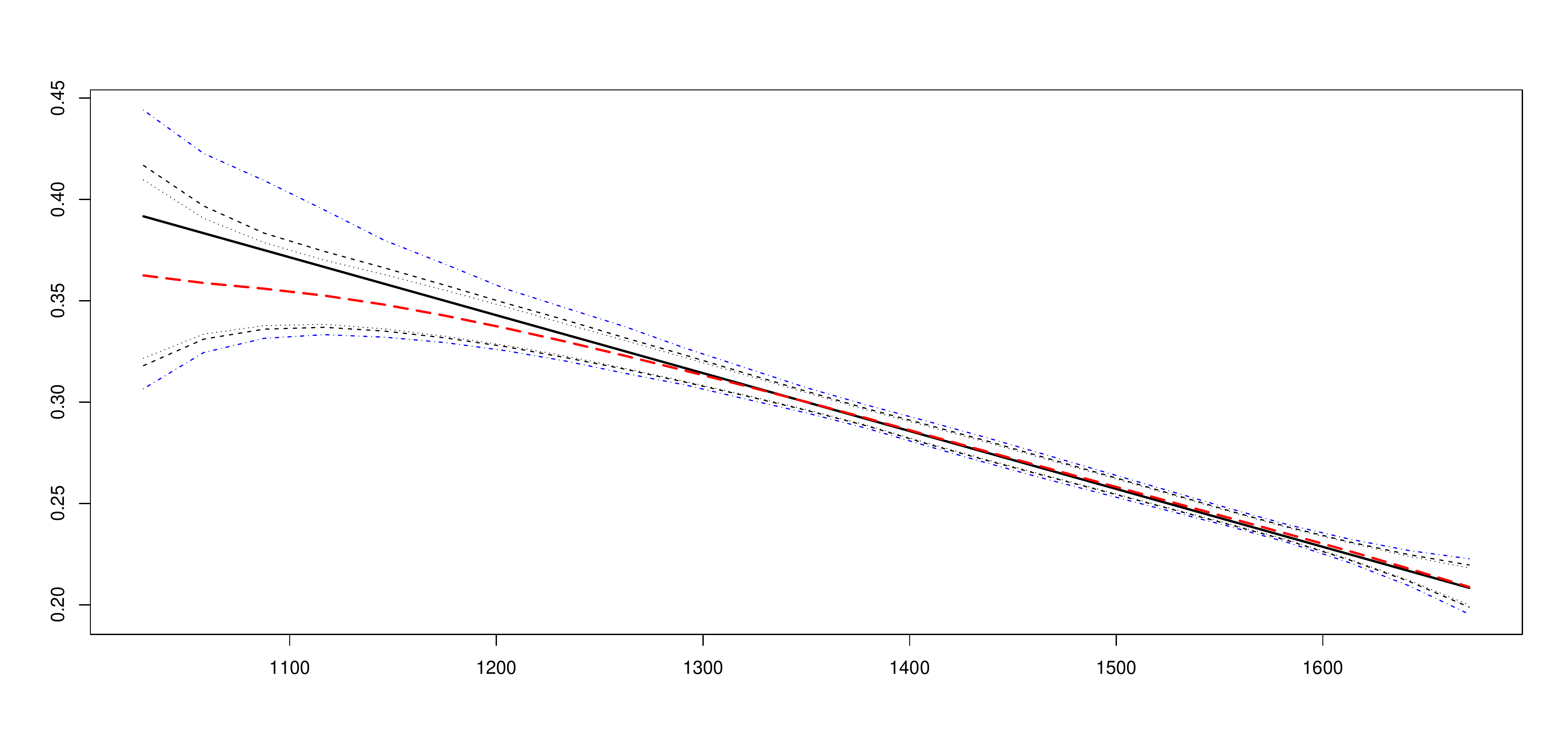}\\

\tiny{Panel C: Probability}\\\vspace{-0.5cm}
\includegraphics[width=\textwidth]{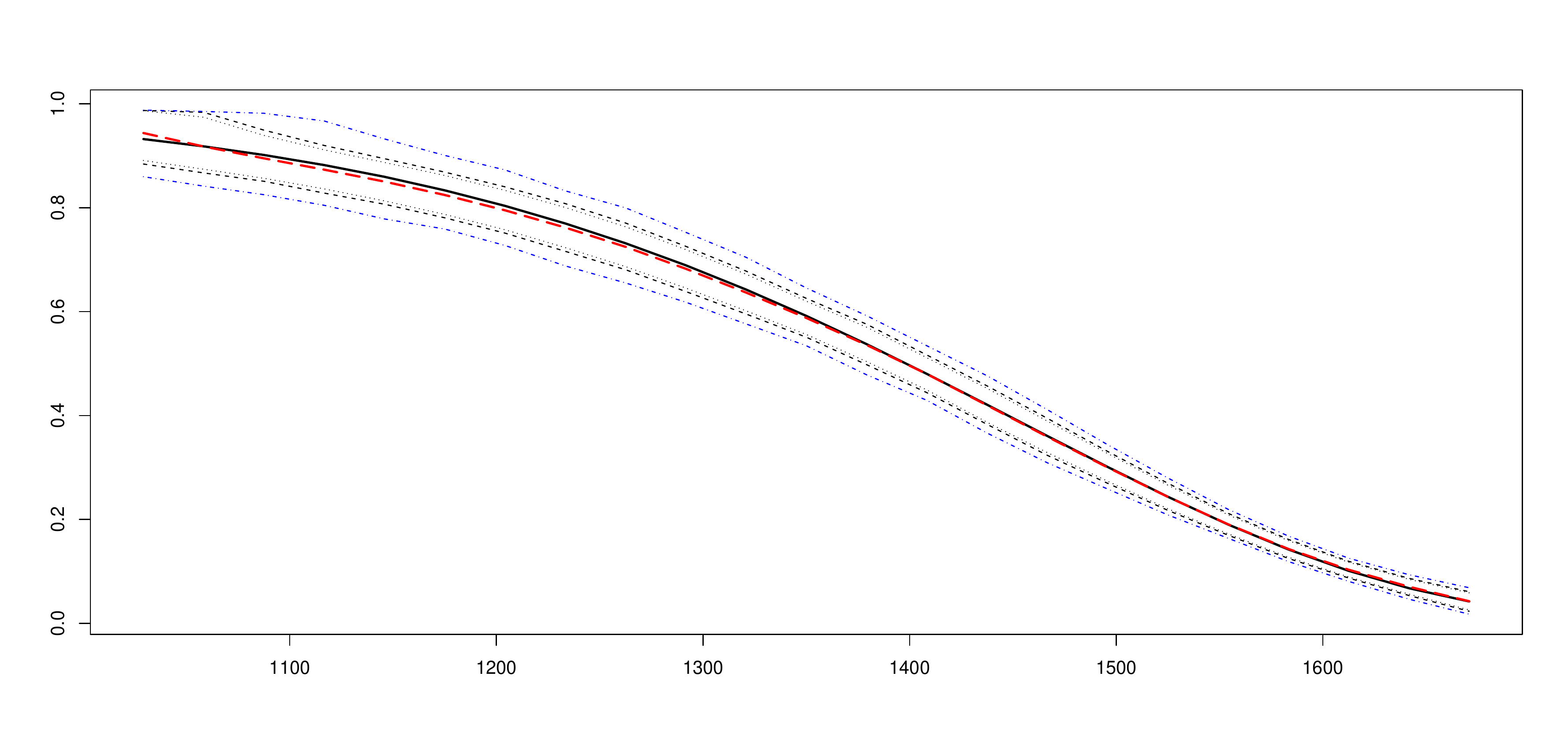}
\caption{\label{fig MC 2}
Simulated Simulated Confidence Intervals: $N=10$ and $\delta=5$.\\
{\tiny
The confidence bands and mean were obtained after $5,000$ Monte Carlo simulations.
}
}
\end{figure}
\end{center}

\begin{center}
\begin{figure}
\tiny{Panel A: Option Price}\\\vspace{-0.5cm}
\includegraphics[width=\textwidth]{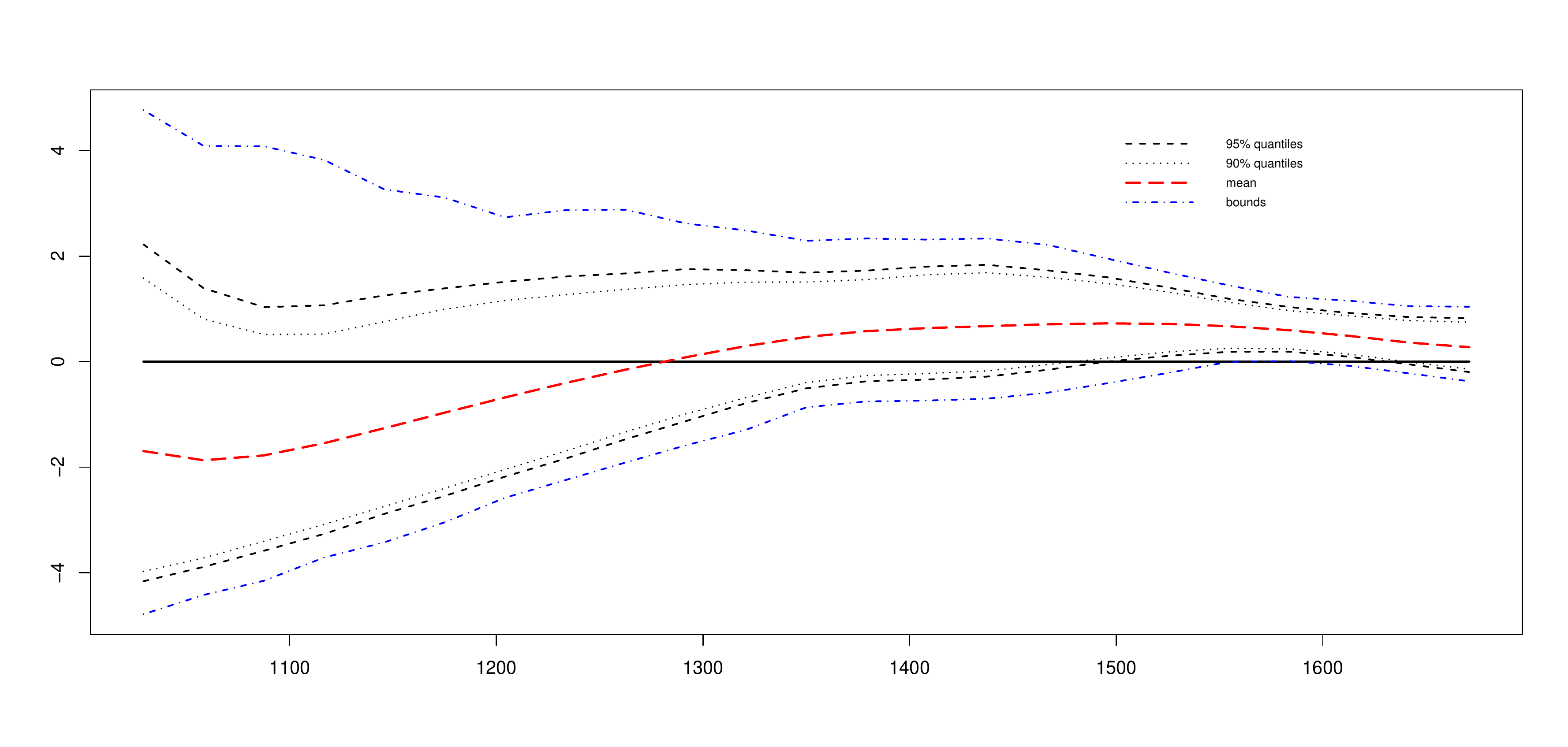}\\

\tiny{Panel B: Volatility}\\\vspace{-0.5cm}
\includegraphics[width=\textwidth]{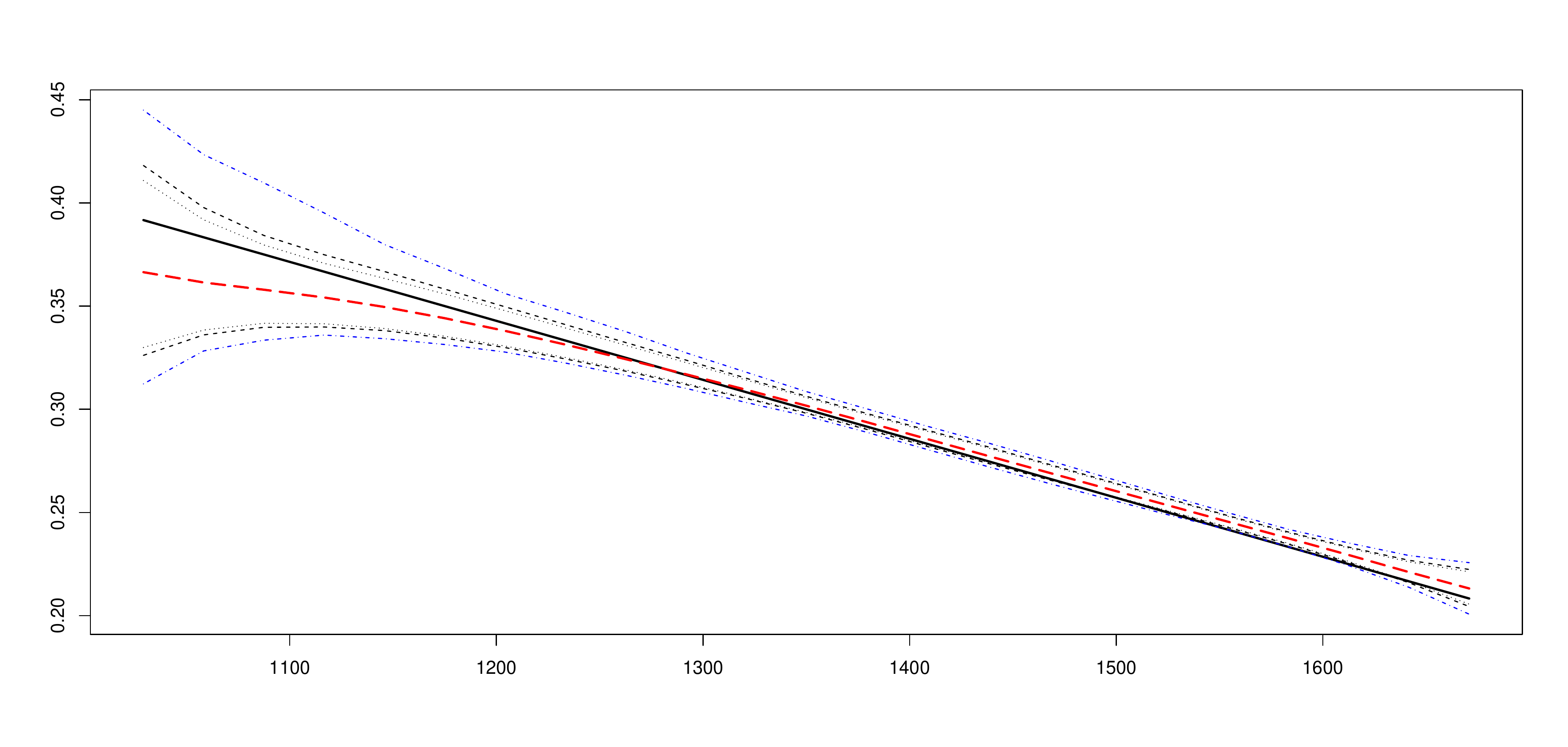}\\

\tiny{Panel C: Probability}\\\vspace{-0.5cm}
\includegraphics[width=\textwidth]{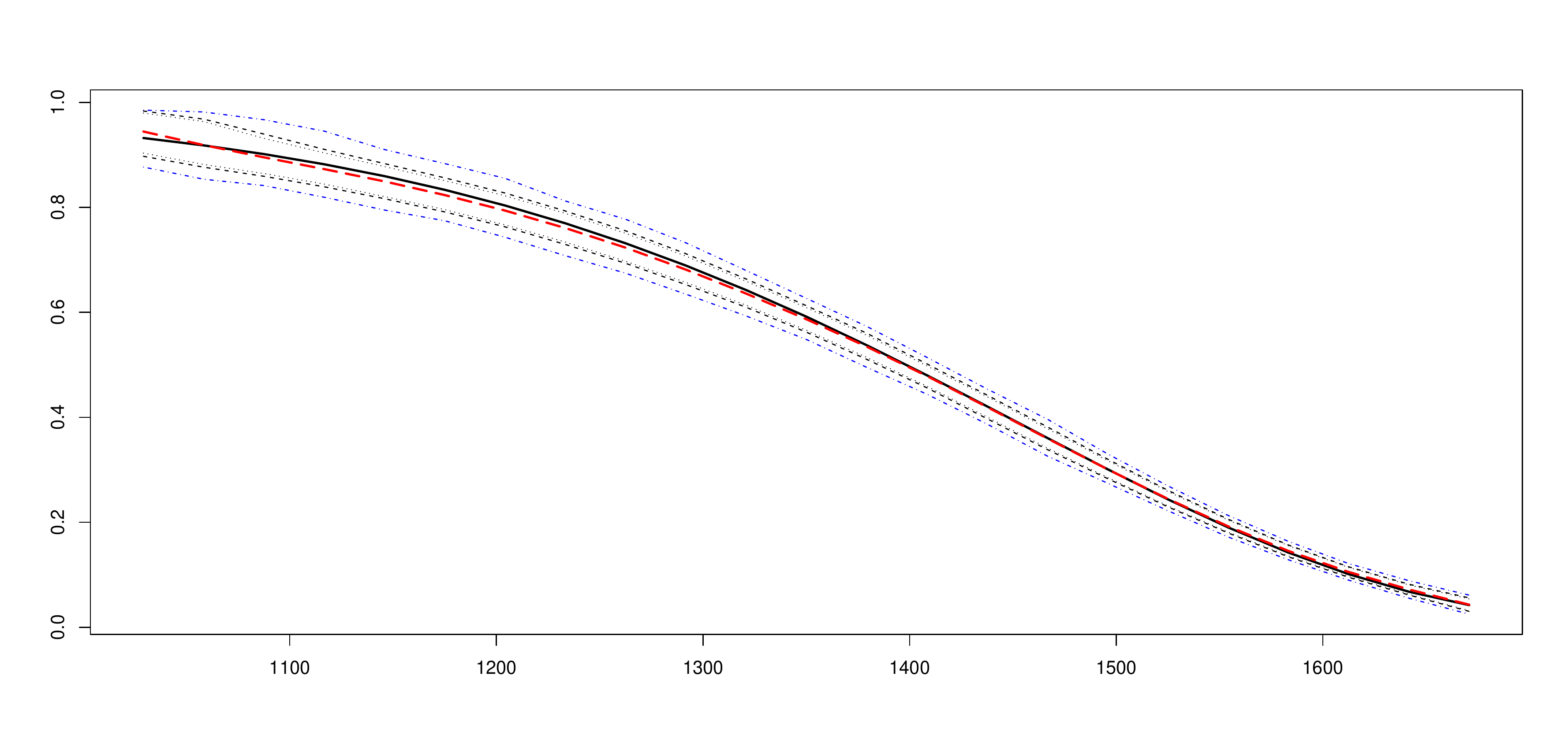}
\caption{\label{fig MC 10}
Simulated Confidence Intervals: $N=10$ and $\delta=10$.\\
{\tiny
The confidence bands and mean were obtained after $5,000$ Monte Carlo 
simulations.
}}
\end{figure}
\end{center}

For any simulation $s=1,\ldots,5000$ we compute, via Theorem 
\ref{th price}, the corresponding CALL function, 
$q^{\delta,s}(k)$, implied volatility , $\sigma^{\delta,s}(k)$, and the associated 
probability $\nu^{\delta,s}$ for different values of $h$, although we only plot 
$\delta=5,10$. For each strike price $k$, we then compute the mean and 
the quantiles of the simulated sample obtaining for each quantity of interest 
maximum and minimum values, $90\%$ and $95\%$ confidence intervals, 
mean and standard deviation. The corresponding curves are plotted in Figure 
\ref{fig MC 2}, for $\delta=5$, and Figure \ref{fig MC 10}, for $\delta=10$.

We notice that, although in the worst possible scenarios the most deeply ITM
options may be overpriced or underpriced by almost $5$ \$ (when the price
fixed by the model is however more than $300$ \$), the mean pricing error
is never larger than $2$ \$ and, for reasonably liquid options just a few cents.
Second, illiquid options are on average underpriced because of the effect
highlighted before by which our method makes mainly use of prices affected 
by negative error terms. This may be considered as an asymmetric smile effect 
resulting, however, from microstructural factors.

%%%

It is noteworthy that for $\delta=5$ the true value always falls inside the 
confidence interval even at the $90\%$ level for all three variables considered, 
suggesting that our approach produces quite reliable predictions. One may 
also notice that the two confidence bounds are not symmetric, especially for 
options deep in the money, reflecting the same distortion noted above. For 
the case $\delta=10$ the situation partly changes as the estimate of the 
option price becomes less precise and, in particular, the mean price exceeds 
the actual one by more than 50 cents for all options ATM or OTM. In addition, 
both lower confidence bounds are breached as frequently as $21.7\%$ 
suggesting that the choice $\delta=10$ produces an increase in smoothness
which results however in a significant pricing error. We investigated also
the results relative to the intermediate values of $\delta$ and we report
the corresponding values of the mean squared error
\begin{equation}
MSE=\frac{1}{5,000\times24}\sum_{k\in K(X)}\sum_{s=1}^{5,000}
\big(q^{\delta,s}(k)-q(k)\big)^2
\end{equation}
which we report in Table 2.

\begin{center}
\vspace{0.5cm}
\begin{tabular}{c|c|c|c|c}
\hline\hline	&$N=4$	&$N=6$	&$N=8$	&$N=10$\\\hline
$\delta=1$	&1.93	&1.99	&2.02	&2.03\\
$\delta=2$	&1.78	&1.87	&1.92	&1.95\\
$\delta=5$	&1.61	&1.63	&1.72	&1.76\\
$\delta=10$	&6.17	&2.06	&1.61	&1.58
\end{tabular}
\vspace{0.5cm}
\nobreak\\\vspace{0.6cm}\nobreak
Table 2. Values of the MSE for alternative parameter choices.
\end{center}

\begin{center}
\begin{figure}
\tiny{Panel A: $\delta=5$}\\\vspace{-0.5cm}
\includegraphics[width=\textwidth]{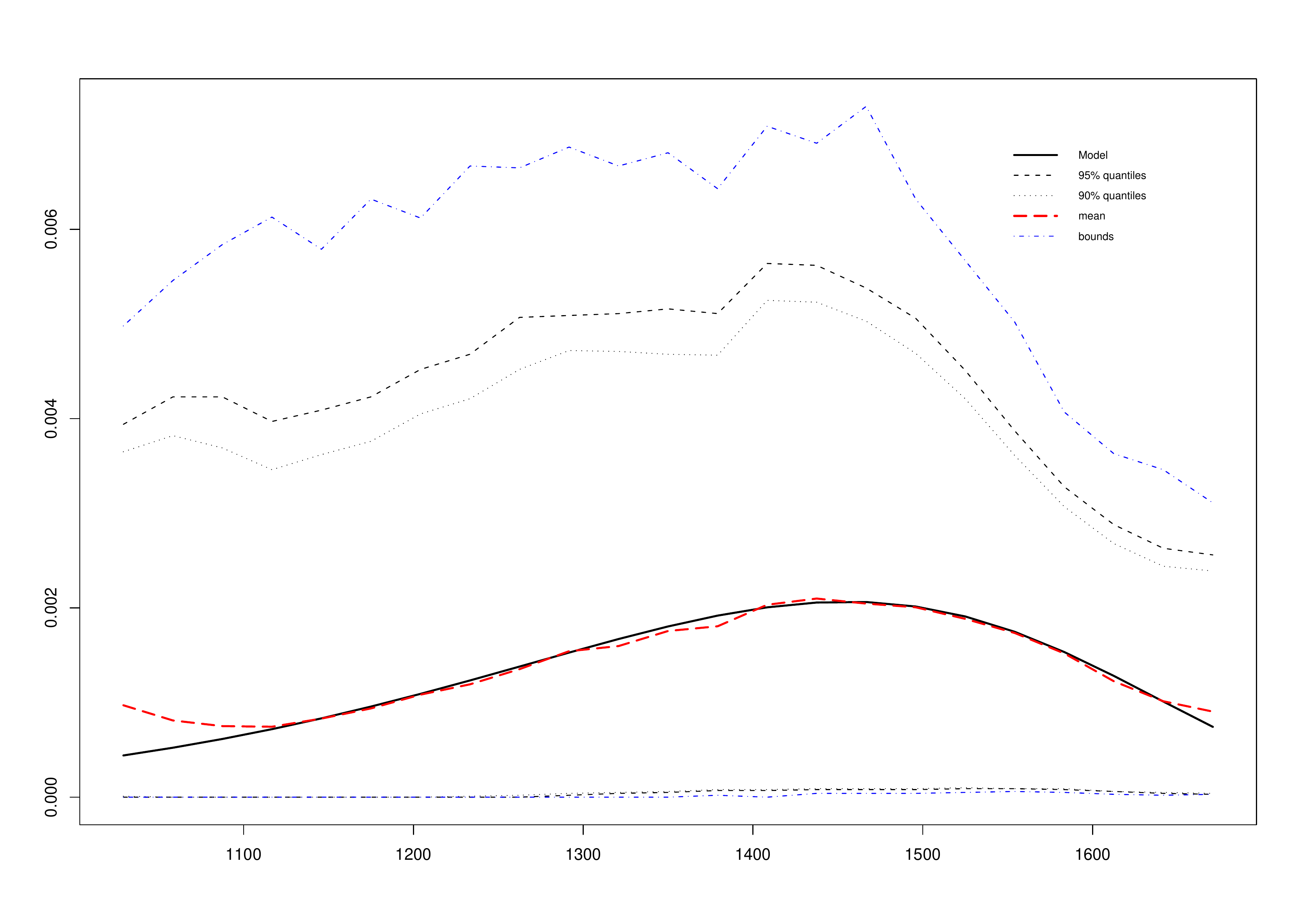}\\

\tiny{Panel B: $\delta=10$}\\\vspace{-0.5cm}
\includegraphics[width=\textwidth]{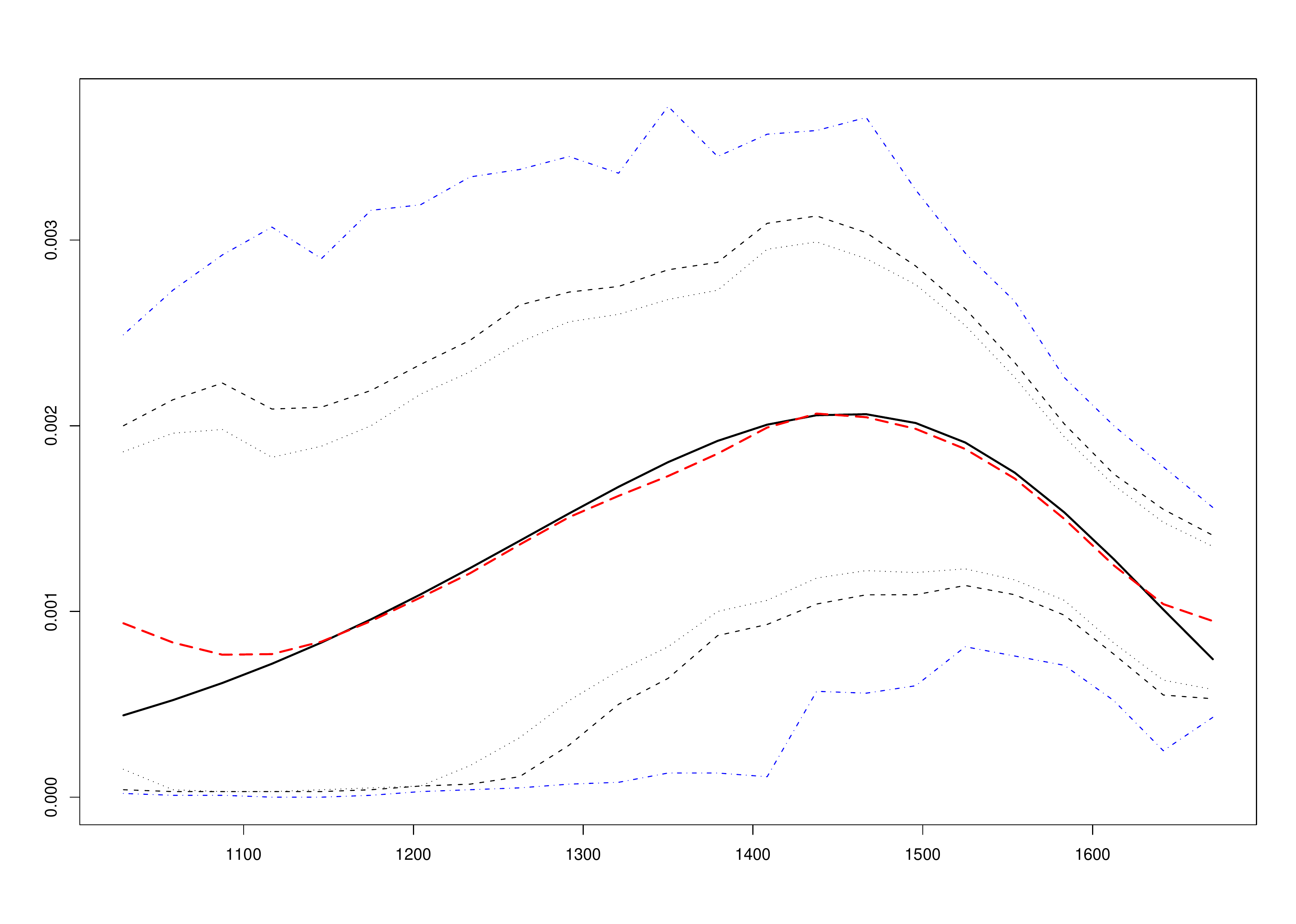}

\caption{\label{fig rnd}
Implied risk-neutral average density: $N=10$ and $\delta=5,10$.
}
\end{figure}
\end{center}

In the literature there is greater emphasis on the implied risk neutral density 
rather than on the implicit probability so that the focus is actually on the 
second derivative of the CALL function. We do not have a special interest
for this quantity here, partly because the presumption that a density actually 
exists has no financial basis. In part, however, it is our choice to work with 
cubic splines that limits our ability to explore densities. Second derivatives in 
fact exist but are piecewise linear, making the candidate density function not 
particularly interesting for applications. To obtain smoothness of the implied
risk neutral density one should perhaps adopt a different functional form than 
cubic splines such as splines of higher order (which are however much less 
tractable computationally speaking) or as the normal option smoother described 
in \eqref{normal}%
\footnote{
We have performed the Monte Carlo analysis described in this subsection also
for option payoffs obtained via \eqref{normal}. Nevertheless the output we
obtained, e.g. in terms of the MSE, is less satisfactory than the results illustrated
here for cubic splines.
}.
Another possibility would be to apply to the density obtained some local
smoothing technique. In the Monte Carlo analysis performed here, however, 
smoothness of the density function arises upon averaging across all simulations. 
In Figure \ref{fig rnd} we plot the estimated mean density function together with 
the one originated from model. The value for the MISE so obtained is $4.402242e-07$
and $4.318549e-07$, for $\delta=5$ and $\delta=10$ respectively.

\subsection{A word on normal option smoothing.}
\begin{center}
\begin{figure}
\tiny{Risk neutral density}\\\vspace{-0.5cm}
\includegraphics[width=\textwidth]{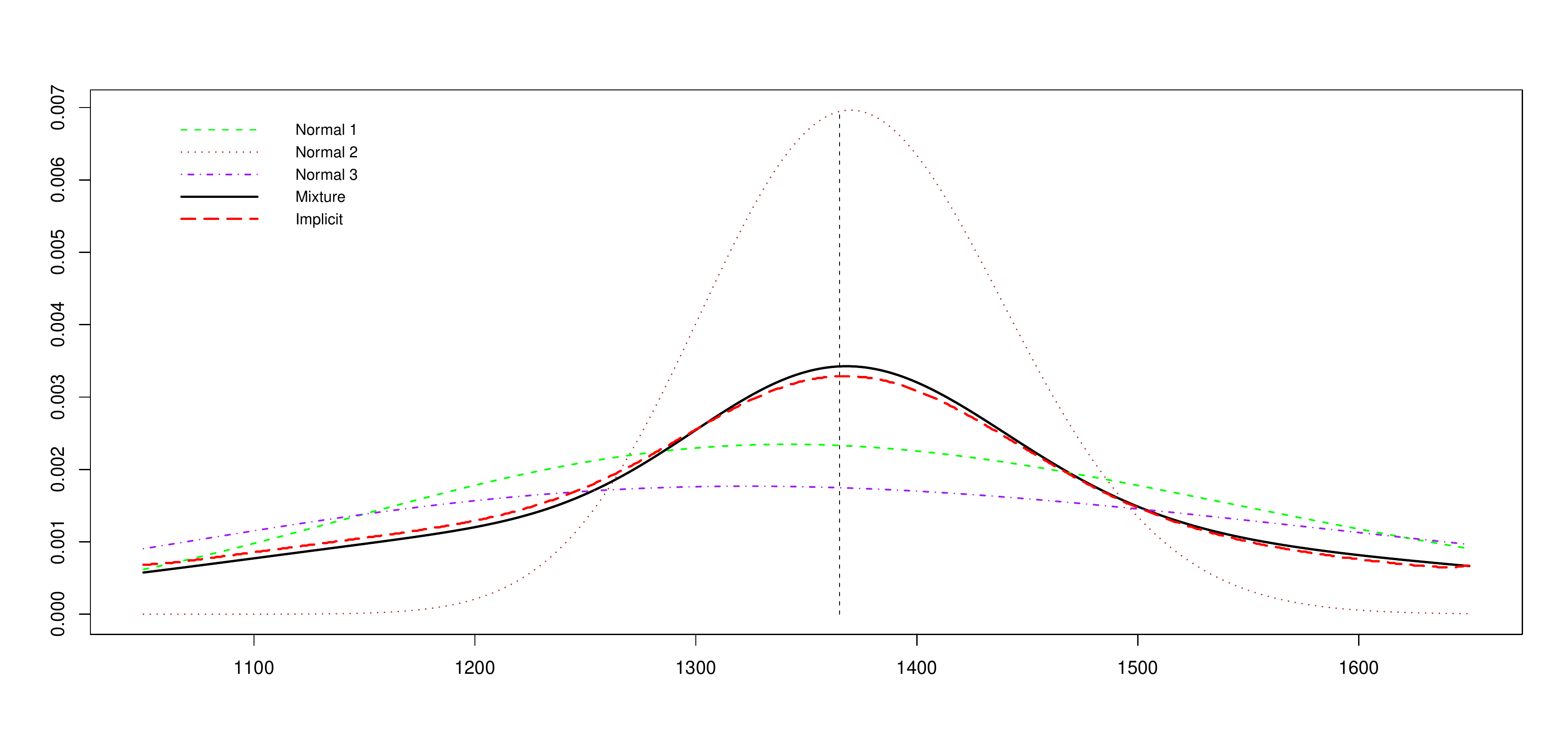}
\caption{\label{fig mixture}
Risk-neutral densities for a mixture of log-normals.\\
{\tiny
All curves represent a corresponding conditional density over the interval
$1050-1650$. The dashed lines are the starting log-normal densities and
the black solid line their mixture. The red line is the density extracted from
the option prices. $MISE=3.34864e-06$.
}
}
\end{figure}
\end{center}

We have performed the above analysis also via the normal options smoothing
formula \eqref{normal} rather than splines. The resulting values for the MSE 
as well as of confidence levels are less satisfactory than those reported above. 
We claim, however, that,in models assuming that the risk neutral density is 
a mixture of log-normals, then the density approach described in \eqref{normal} 
produces remarkable results. To this end we have considered three normal 
densities with instantaneous parameters $\mu=\{0.027,0.033,0.049\}$ and 
$\sigma=\{0.3,0.1,0.4\}$ and weights $\alpha=\{0.2,0.3,0.5\}$, respectively.
We have used the same values for strike prices, underlying, maturity and interest 
rate as above. Given these values%
\footnote{
In fact with the above values, $\sum_{i=1}^r\alpha_i\mu_i=r$.
},
we obtain risk neutral prices for options with strike prices ranging from 1,000 to 
1,700. From such prices we extract the implied risk neutral density by using 
\eqref{normal} after setting $h=20$. We plot in Figure \ref{fig mixture} the 
conditional risk-neutral densities over the interval $1,050-1,650$. We stress 
that indeed the mixture of log-normals is rather fat tailed, as desired, but 
also that its non parametric estimate is indeed very close to it. The corresponding 
value for the MISE amounts to $3.34864e-06$.

\section{Empirical Applications: Market Data}
\label{sec market}

\begin{center}
\begin{figure}
\includegraphics[width=1\textwidth]{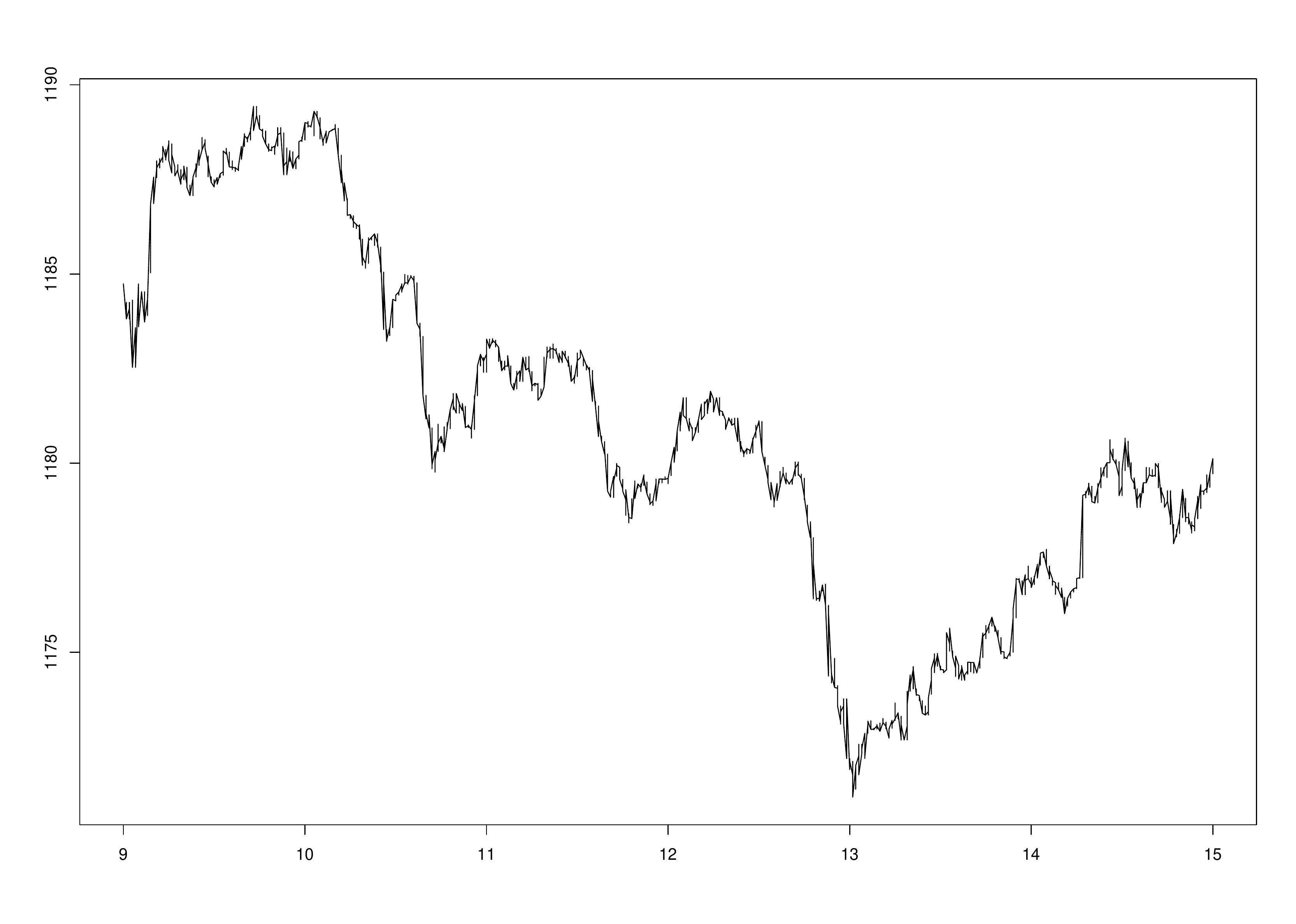}
\caption{The S\&P 500 Index on 21$^{st}$ October 2010}
\label{fig SP}
\end{figure}
\end{center}

Eventually, we consider an application to market data by selecting an arbitrary 
trading day, 21$^{st}$ October 2010, on the S\&P 500 options market%
\footnote{
We make use of the quote prices provided by CBOE Market Data Retrieval (MDR). 
The dataset contains, among other things, information on bid and ask prices and 
volumes. Data are sampled at a frequency higher than 1 minute.
}. 
We 
sample ask quotes at time intervals of one minute each and disregard quotes 
for which the reported ask size is below 100. On the subsample so obtained 
we have options quotes for 180 different strike prices -- ranging from 50 to 2500 
-- and 12 possible maturities -- from $22^{nd}$ October 2010 to $22^{nd}$ 
December 2012. We focus on options expiring in November 2010 and their 
quotes at 12:06, 12:39 and 13:03, around the market downturn of Figure 
\ref{fig SP}.

At the three selected times and for the selected maturity there are 65, 53 and 
67 efficiently quoted strike prices respectively out of 90, 75 and 99. We therefore 
have a relatively long cross section of strikes and an incidence of inefficient 
prices of $30\%$ on average.  The large number of available strikes is one of the 
advantages of working with quoted ask prices, as dictated by our model, rather 
than transaction prices. In the sample there is an overwhelming ratio of 
contracts ITM by $10\%$ or more and virtually no OTM contract  as reported in 
Table 3.
\begin{center}
\vspace{0.5cm}
\begin{tabular}{r|rrrrcc}
\hline\hline&Total	&dITM	&ITM		&ATM	&OTM	&dOTM\\\hline
Sample	&23.972	&45.53	&27.15	&27.22	&0.09	&0\\
12:06	&90		&39.77	&35.23	&25.00	&0		&0\\
12:39	&75		&52.31	&32.31	&15.38	&0		&0\\
13:03	&99		&38.37	&31.39	&30.23	&0		&0\\\hline
\end{tabular}
\nobreak\\\vspace{0.6cm}\nobreak
Table 3: Contracts by Moneyness, $S_t/K$.
\vspace{0.5cm}
\end{center}

We select a subsample of strikes ranging from 670 to 1,255. At 12:39 the lesser 
number of strikes quoted corresponds to a larger maximum interval between 
consecutive strikes, i.e. $M=70$, while at the other moments strikes do not differ 
by more than 25 and 30, respectively. Thus for the typical choice $\delta=5$ we 
expect to have a relatively poor performance at $12:39$, due to the high value
of $M$. In the following picture we plot for each time $t$ the curves corresponding 
to the three distinct values of $\delta=2,5,10$.

\begin{center}
\begin{figure}
\tiny{Panel A: Option Prices at 12:06}\\\vspace{-0.7cm}
\includegraphics[width=0.65\textwidth]{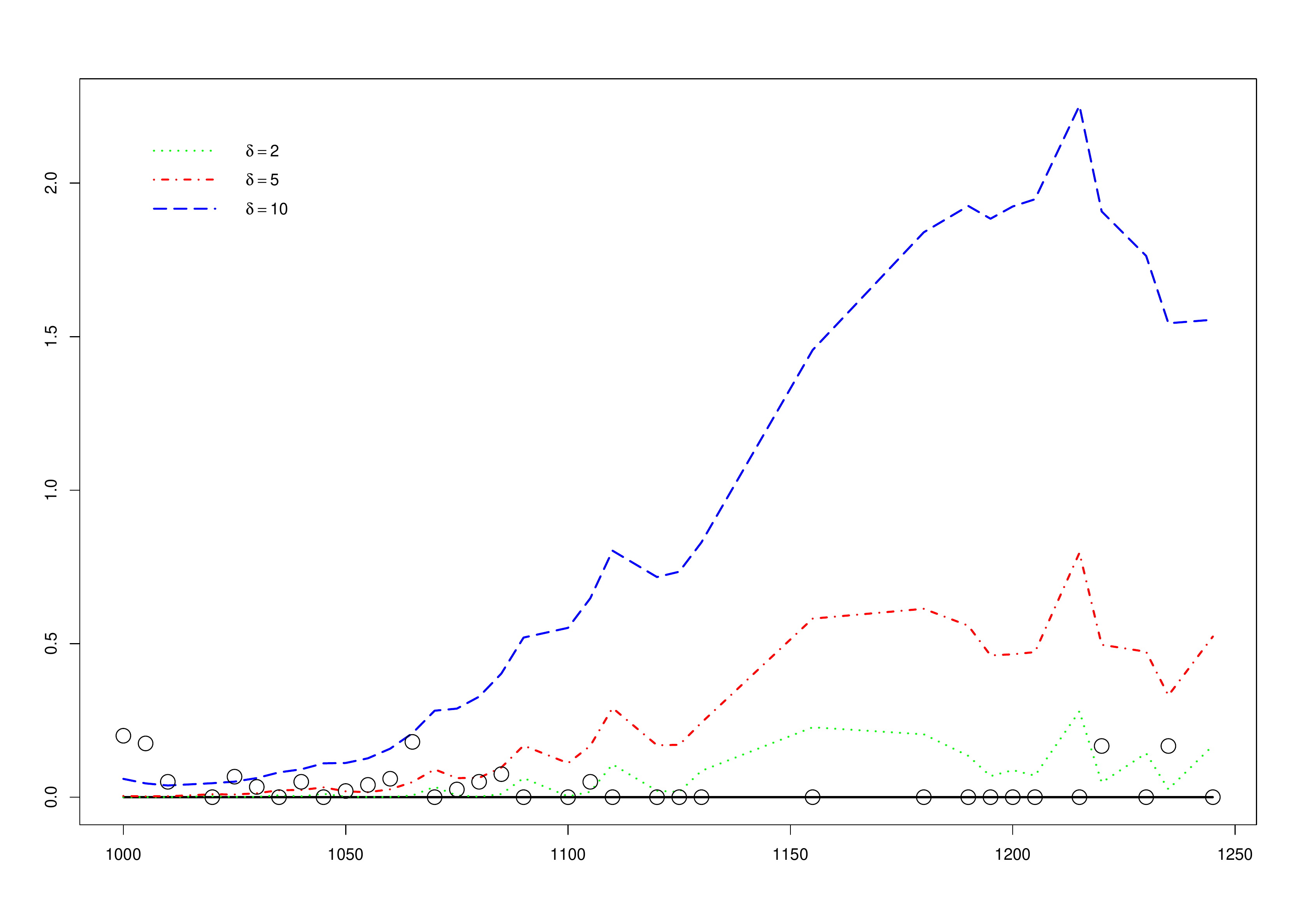}\\
\tiny{Panel B: Option Prices at 12:39}\\\vspace{-0.7cm}
\includegraphics[width=0.65\textwidth]{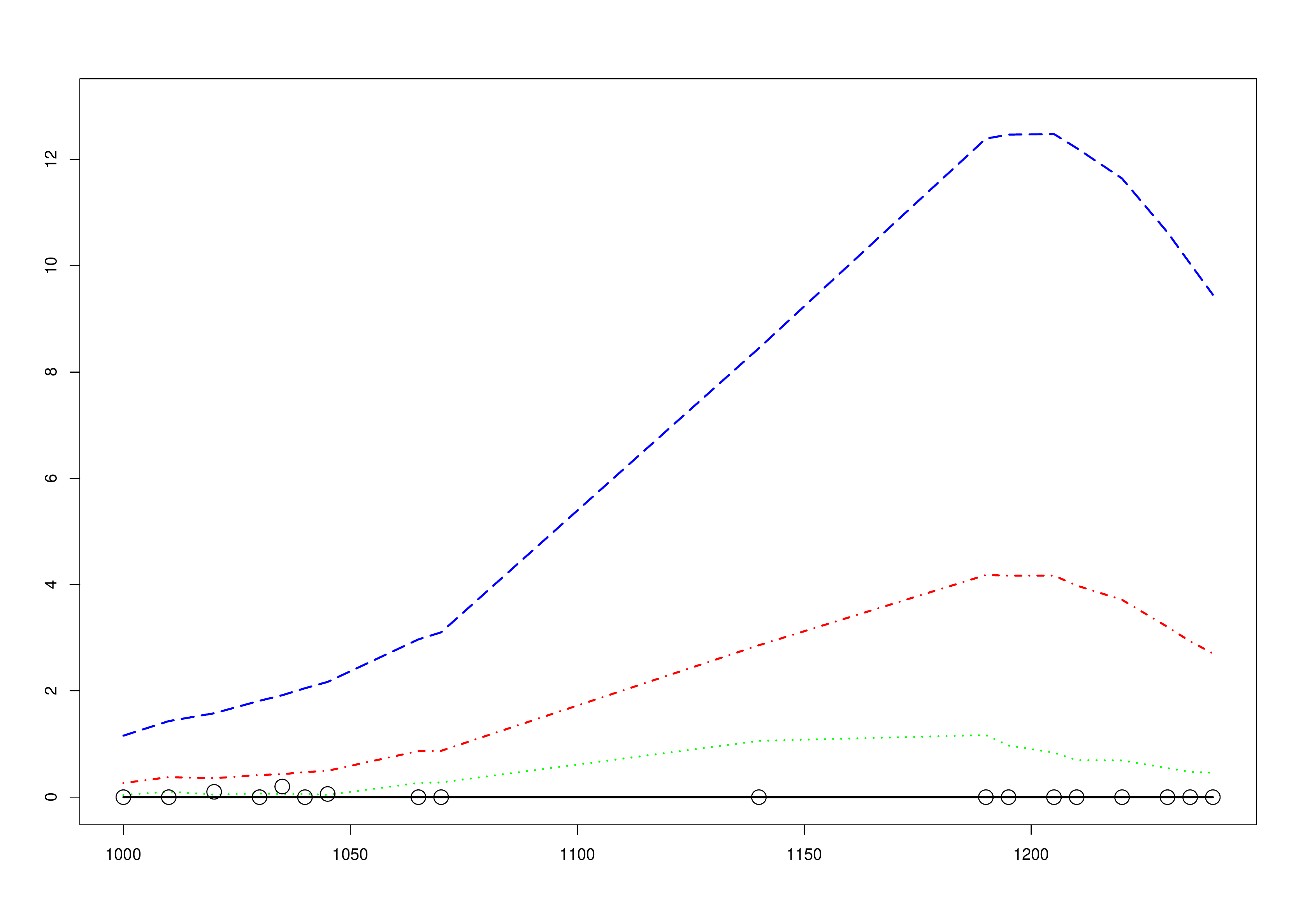}\\
\tiny{Panel C: Option Prices at 13:03}\\\vspace{-0.7cm}
\includegraphics[width=0.65\textwidth]{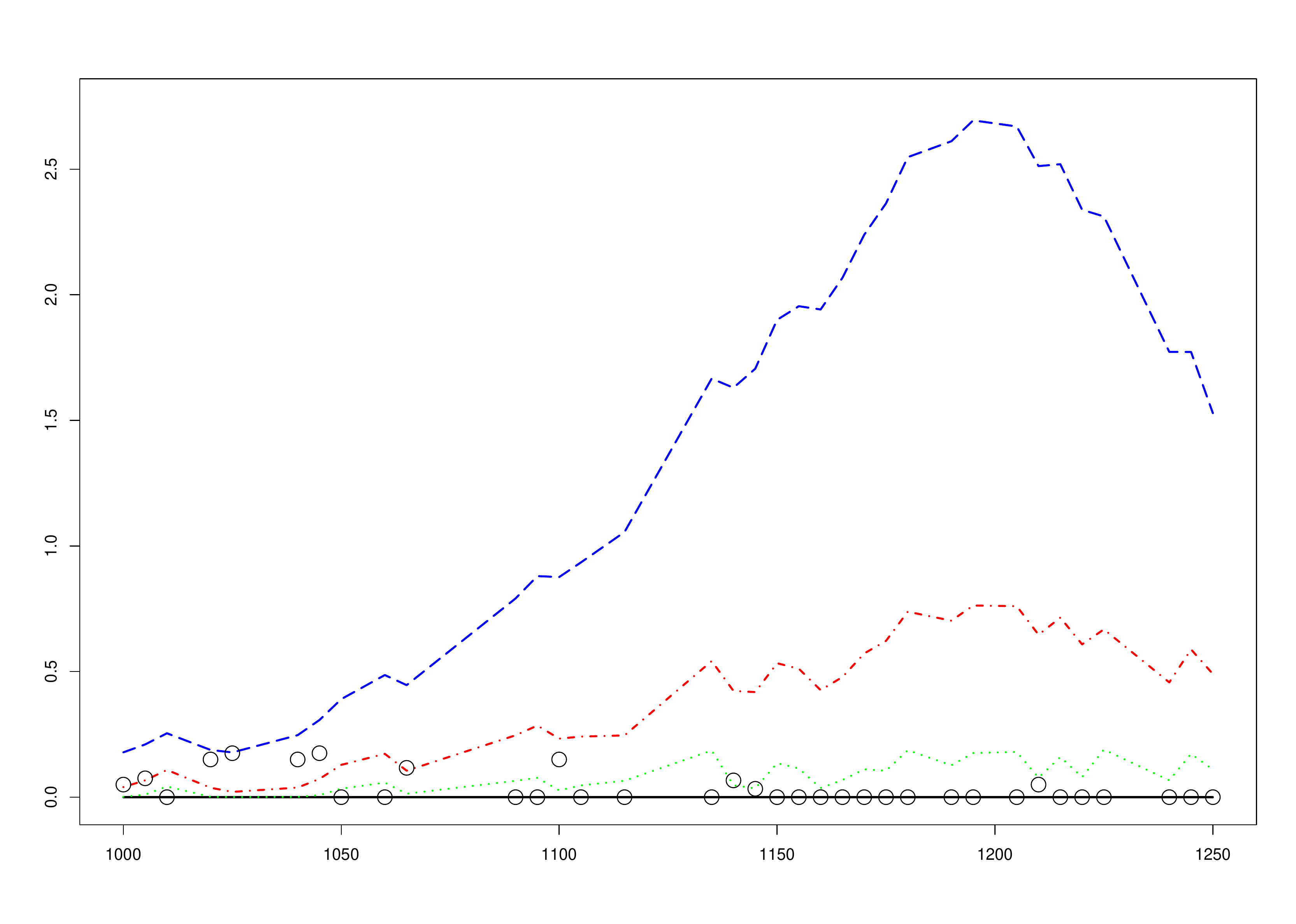}
\caption{
Actual and estimated option prices as differences with the
efficient prices. Market prices are represented as circles.
\label{fig estimated q}}
\end{figure}
\end{center}

In fact we clearly see from Figure \ref{fig estimated q} that the performance
of our estimates at $12:39$ is quite poor due to the fact that there is just
one quoted strike between 1075 and 1175. The price curve ends up being
very smooth but overestimates actual prices by as much as $5\$$ for the
value $\delta=5$ while in the other selected instants the price gap never 
exceeds $50$ cents for such a parameter choice.

\begin{center}
\begin{figure}
\tiny{Panel A: Implied risk neutral probability}\\\vspace{-0.7cm}
\includegraphics[width=0.8\textwidth]{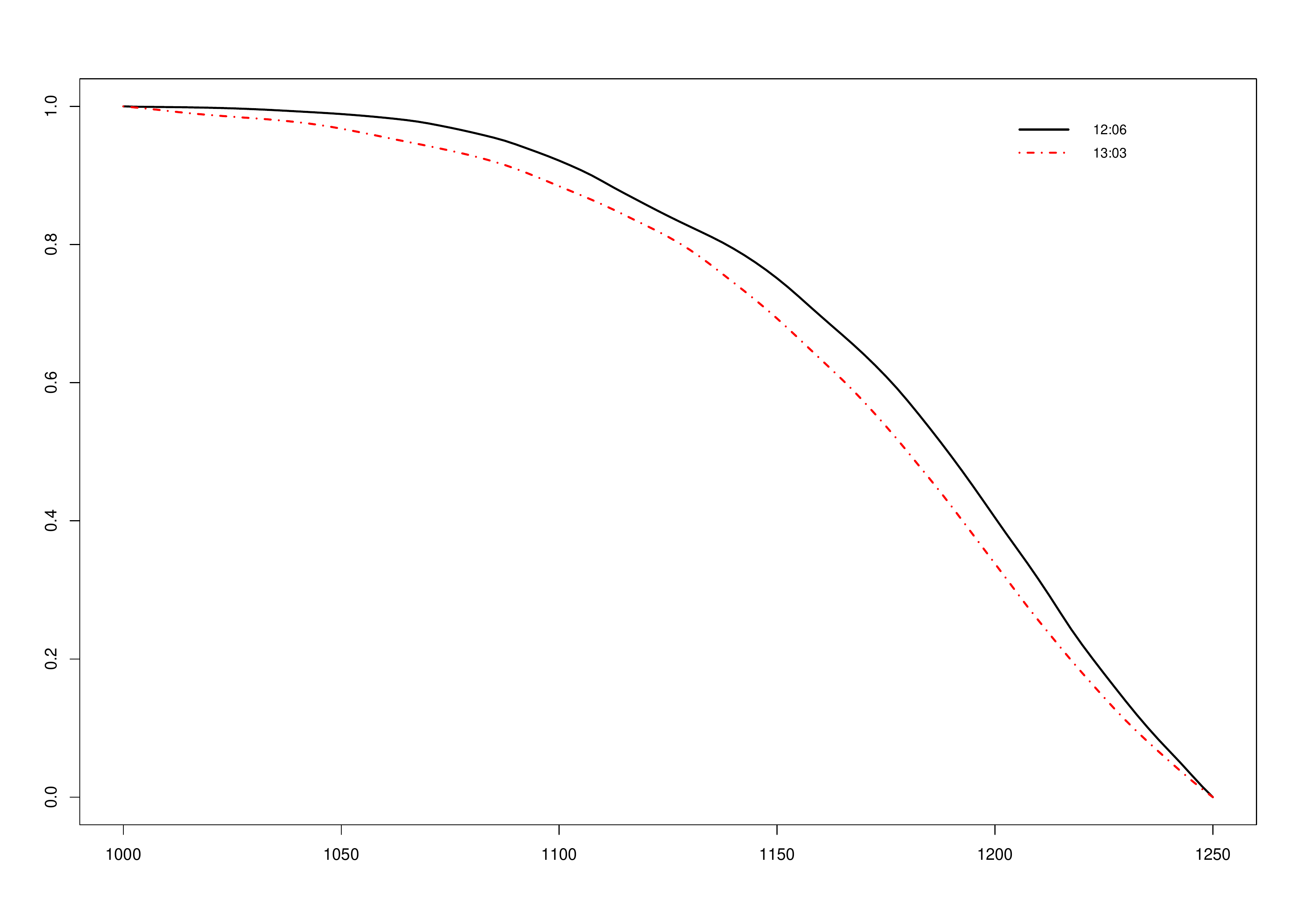}\\
\tiny{Panel B: Implied risk neutral density}\\\vspace{-0.7cm}
\includegraphics[width=0.8\textwidth]{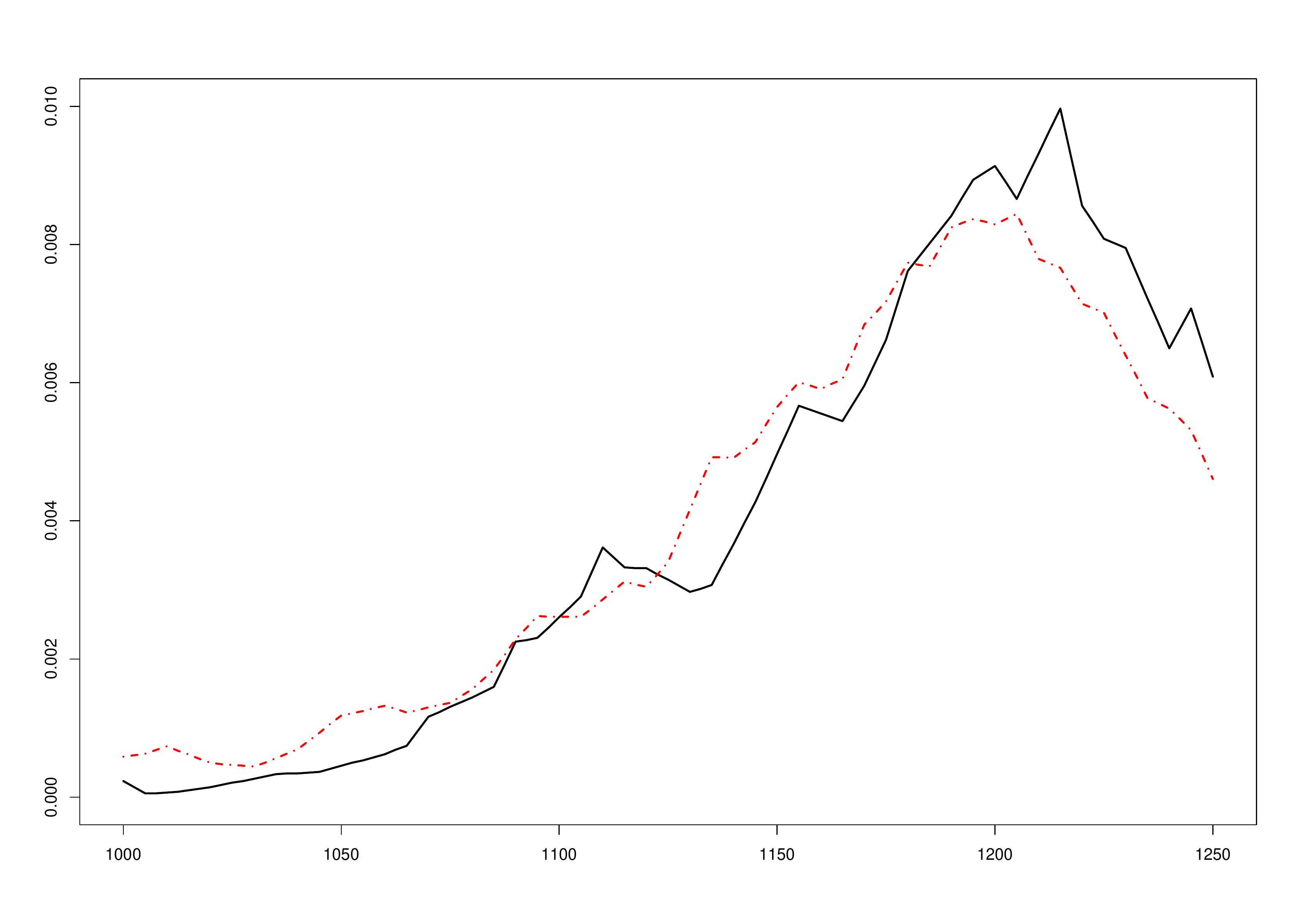}
\caption{
Conditional risk neutral distribution, $\nu^\delta(X>x)$, and density, 
$d^\delta(x)$, at 12:06 and 13:03 
for the case $\delta=5$.
\label{fig Nu Compare}}
\end{figure}
\end{center}

Eventually we plot, for the value $\delta=5$, the implied conditional risk 
neutral probability and density at $12:06$ and $13:03$, Figure \ref{fig Nu Compare}, 
to capture the effect of the market downturn on $\nu^\delta$ and $d^\delta$. 
As expected, the fall in the underlying price determines a more pessimistic view 
embodied in the implied risk neutral distribution. Of course, as explained above, 
the density is not a smooth quantity, due to our choice of working with cubic splines.

Eventually, we can use the above information to compute some measures 
of risk. In particular we consider a long position in a future contract 
expiring on November 2010, i.e. in a month. The future price is set
according to the future/spot parity, $F(t,T)=S_t\exp(r(T-t))$. As a
proxy for the riskless rate $r$ we use the 1 month LIBOR rate on that date, 
quoted at $3.96\%$. We then compute the $VaR$ and conditional $VaR$ at 
a confidence level of $2\%$ and $5\%$ at the two instants of time. The
values are reported in the following Table 4:

\begin{center}
\begin{tabular}{l|r|r}
\hline\hline
			&2.5\%		&5\%\\\hline
$VaR$  12:06	&177.58		&110.08\\
$VaR$  13:03	&163.94		&123.95\\
$CVaR$ 12:06	&183.00		&134.30\\
$CVaR$ 13:03	& 171.26		&149.65\\\hline
\end{tabular}
\nobreak\\\vspace{0.6cm}\nobreak
Table 4: Measures of risk.
\vspace{0.5cm}
\end{center}

These last remarks suggest the importance to investigate the dependence of $\nu$ 
on the current value of the underlying, although outside of our interests here. Another
issue that would be important to address is the time evolution of the pricing measure.

\appendix

\section*{Mathematical Appendix}

In this appendix we present some results which we used in the proofs of the main
Theorems.

\section*{Results from section \ref{sec option}}

\begin{proof}[\textbf{Proof of Theorem \ref{th price}}]
The claim essentially follows from \cite[Lemmas 7 and 8]{ThOpt} in which, 
however, it is assumed that $x<\infty$. The proof given here is adapted from
that one to cover the present setting. First of all, in superhedging a given
claim, we can restrict to the options which are priced efficiently, i.e. whose
strike is included in $K^0(X)$. Let $j_0=0$ and write $j_{I+1}=x$, if $x<\infty$ 
or else $j_{I+1}>j_I$ arbitrarily. $g^0_\theta(X)\ge f(X)$ for $X=0$. Assume 
that $g_\theta^0(j_i)<f(j_i)-\varepsilon$ for some $i=1,\ldots,I+1$. By continuity 
there exists then $\delta>0$ such that $g^0_\theta(X)<f(X)-\varepsilon$ holds 
on the set $\{j_i-\delta<X\le j_i\}$. However, $P(j_i-\delta<X\le j_i)>0$ since 
otherwise $\ess\sup(X\wedge j_i)\le j_i-\delta$ in contrast with Assumption 
\ref{ass X}. Thus $g^0_\theta(X)\ge f(X)$ a.s. implies that 
$g^0_\theta(j_i)\ge f(j_i)$ for $i=0,\ldots,I+1$. If $\bar X=\infty$ then by the 
fact that $j_{I+1}$ was chosen arbitrarily we deduce $g^0_\theta(x)/x\ge f(x)/x$, 
i.e. $1\ge \hat f(\bar X)$.

Viceversa, if $g_\theta^0(j_i)\ge f(j_i)$ for $j=0,\ldots,I+1$, then for each 
$\omega\in\{X\le\bar X\}$ and some choice of $j_{I+1}$ there exists
$i(\omega)\in\{0,1,\ldots,I\}$ such that, for some $0\le a\le1$,
$j_{i(\omega)}\le X(\omega)\le j_{i(\omega)+1}$ and thus
\begin{align*}
f(X(\omega))
	\le
af(j_{i(\omega)})+(1-a)f(j_{i(\omega)+1})
	\le
ag_\theta^0(j_{i(\omega)})+(1-a)g_\theta^0(j_{i(\omega)+1})
	=
g^0_\theta(X(\omega))
\end{align*}
because $f$ is convex and $g^0_\theta$ is linear on each interval $[j_i,j_{i+1}]$. 
In other words, $g^0_\theta(X)\ge f(X)$ outside of $\{X>\bar X\}$ i.e. $P$-a.s.
so that the inequality $g^0_\theta(X)\ge f(X)$ $P$-a.s. may be written in vector
notation as 
\begin{align*}
\mathbf D\mathbf a\ge\mathbf f
\end{align*}
where $\mathbf a$ is the vector of weights $a_1,\ldots,a_I$ such that
$\theta=\sum_{i=1}^Ia_i\theta(j_i)\in\Theta$, $\mathbf D$ is the 
$(I+1)\times(I+1)$ matrix whose inverse appears in \eqref{qg} and
$\mathbf f=[f(j_1),\ldots,f(j_I),\hat f(\bar X)]^T$. 

Define the vectors $\mathbf w,\mathbf b\in\R^{I+1}$ implicitly by letting
\begin{equation}
\label{b recursion}
b_Id_I
	=
q(j_I)
\quad\text{and}\quad
b_Ie_I+\sum_{i=n}^{I-1}b_i
	=
\frac{q(j_n)-q(j_{n+1})}{j_{n+1}-j_n}
\qquad n=0,\ldots,I-1
\end{equation}
with $e_I=1$ if $\bar X<\infty$ or else $e_I=0$ and
\begin{equation}
\label{w}
\sum_{i=0}^nw_i
	=
\frac{f(j_{n+1})-f(j_n)}{j_{n+1}-j_n}
\qquad
n=0,\ldots,I-1
\quad\text{and}\quad
\sum_{i=0}^Iw_id_i=\hat f(\bar X)
\end{equation}
Clearly, $w$ and $b$ satisfy \eqref{wb}.
The following properties are easily established by induction: 
(\textit i) $\mathbf b\ge0$ (as $j_0,\ldots,j_I\in K^0(X)$), 
(\textit{ii}) $\mathbf w\ge0$ (as $f\in\Gamma$), 
(\textit{iii}) $\mathbf b^T\mathbf D=\mathbf q^T$ and 
(\textit{iv}) $\mathbf w=\mathbf D^{-1}\mathbf f$. But then,
\begin{align*}
q(\theta(f))
	&=
\min_{\left\{\mathbf a\in\R^{I+1}_+:\mathbf D\mathbf a\ge\mathbf f\right\}}%
\mathbf q^T\mathbf a
	=
\min_{\left\{\mathbf a\in\R^{I+1}_+:\mathbf D\mathbf a\ge\mathbf f\right\}}%
\mathbf b^T\mathbf D\mathbf a
	\ge
\mathbf b^T\mathbf f
	=
\mathbf q^T\mathbf w
\end{align*}
\end{proof}

\section*{Results from section \ref{sec estimate}}

\begin{proof}[\textbf{Proof of Lemma \ref{lemma turlach}.}]
By a result of Turlach \cite[p. 85]{turlach} the program \eqref{spline} 
admits as its solution a cubic $\mathscr C^2$ spline of the form
\begin{equation*}
g_k^h(x)
	=
\sum_{i=1}^N\set{\big[k+t^h_i,k+t^h_{i+1}\big)}(x)\Phi_{k,i}^h\Big(x-(k+t^h_ i)\Big)
	+
\set{\big[k+t^h_I,\infty\big)}(x)(x-k)
\qquad
x\in\R_+
\end{equation*}
where $t_1^h,\ldots,t_{N+1}^h$ are as in the text and $\Phi^h_{k,i}$ is a 
polynomial of degree $3$ for $i=1,\ldots,I$. It is clear from the constraints 
imposed to \eqref{spline} that indeed $g_k^h\in\Gamma$. Moreover, these 
same constraints imply that $g_k^h(x)=g^0_k(x)$ when $x\notin[k-h,k+h]$ 
while $1\ge Dg_k^h(x)\ge0$ on $[k-h,k+h]$. Thus, $g_k^h-g^0_k$ is increasing 
on $(-\infty,k]$ and decreasing afterwards, so that
$
\sup_x(g^h_k-g^0_k)(x)
\le
(g^h_k-g^0_k)(k)=g^h_k(k)
$. 
Moreover, the (non empty) set of solutions is clearly convex and the functional
$I_k(h;f)$ is strictly convex in $f$ so that the solution is necessarily unique.
Define
\begin{equation*}
\bar g_k^h(x)=g_k^h(2k-x)+(x-k)
\qquad x\ge0
\end{equation*}
and observe that $\bar g_k^h\in\chi(k;h)$. Moreover, one deduces from \eqref{I} that
\begin{align*}
I_k(h;\bar g_k^h)
	&=
\sum_{i=1}^{N+1}\big[g_k^0(k+t_i^h)-\bar g_k^h(k+t_i^h)\big]^2
	+
(0.1h)^3\int_{k-h}^{k+h}\big(D^2\bar g_k^h(x)\big)^2dx\\
	&=
\sum_{t_i^h\le0}\big[\bar g_k^h(k+t_i^h)\big]^2
	+
\sum_{t_i^h>0}\big[\bar g_k^h(k+t_i^h)-t_i^h\big]^2
	+
(0.1h)^3\int_{k-h}^{k+h}\big(D^2\bar g_k^h(x)\big)^2dx\\
	&=
\sum_{t_i^h\ge0}\big[g_k^h(k+t_i^h)-t_i^h\big]^2
	+
\sum_{t_i^h<0}\big[g_k^h(k+t_i^h)\big]^2
	+
(0.1h)^3\int_{k-h}^{k+h}\big(D^2g_k^h(x)\big)^2dx\\
	&=
I_k(h;g_k^h)
\end{align*}
However, since the solution is unique, we have the symmetry relation
\begin{equation}
\label{symmetry}
g_k^h(x)=g_k^h(2k-x)+(x-k)
\qquad x\ge0
\end{equation}
Let $f\in\chi(k;h)$, $h>h'$ and define $T:\chi(k;h)\to\chi(k;h')$ implicitly
by letting
\begin{equation*}
Tf(x)
=
f\left(k+(x-k)\frac{h}{h'}\right)\frac{h'}{h}
\end{equation*}
Observe that $T$ is one to one and onto and that 
$D^2Tf(x)=h/h'D^2f(k+(x-k)h/h')$. Thus,
\begin{align*}
I_k(h';Tf)
	&=
\sum_{i=1}^{N+1}\Big[g_k^0\big(k+t_i^{h'}\big)-Tf\big(k+t_i^{h'}\big)\Big]^2
+
(0.1h')^3\int_{k-h'}^{k+h'}\big(D^2Tf(x)\big)^2dx\\
	&=
(h'/h)^2\sum_{i=1}^{N+1}\Big[g_k^0\big(k+t_i^h\big)-f\big(k+t_i^h\big)\Big]^2
+
(0.1 h')^3(h/h')\int_{k-h}^{k+h}\big(D^2f(z)\big)^2dz\\
	&=
(h'/h)^2I_k(h;f)
\end{align*}
Thus $f$ solves the program \eqref{spline} relatively to $h$ if and only if 
$Tf$ solves it relatively to $h'$. By uniqueness we conclude that 
\eqref{spline projection} holds. Let $0<h'<h$ and observe that, if $x\le k$
\begin{align*}
g_k^{h'}(x)
	=
g_k^h(k+(x-k)h/h')h'/h
\le
g_k^h(x)h'/h
\le
g_k^h(x)
\end{align*}
a conclusion which extends to $x>k$ by \eqref{symmetry}. This proves (\textit{ii}).
Given that $0\le Dg_k^h(x)\le1$ we conclude that 
\begin{align*}
0
	\le
g_k^{h'}(x)-g_k^0(x)
	\le
g_k^{h'}(k)
	=
\frac{h'}{h}g_k^h(k)
\end{align*}
so that $g_k^h$ decreases to $g_k^0$ uniformly in $x$.

If $y=k'-k\in\R$, then writing $t^h_{I+2}=\infty$ and
\begin{align*}
g_k^h(x)
	&=
\sum_{i=1}^{N+1}\set{\big[k'+t^h_i,k'+t^h_{i+1}\big)}(x+y)\Phi_{k,i}^h\big(x+y-(k'+t^h_ i)\big)
	\equiv
\gamma_k^h(x+y;k')
\end{align*}
It is obvious that $\gamma_k^h(x+y;k')=\gamma_k^h(x;k)$ and that 
$g^0_{k+y}(x+y)=g^0_k(x)$. But then, for $i=1,\ldots,N+1$,
\begin{align*}
g^0_{k'}(k'+t^h_i)-\gamma_k^h(k'+t^h_i;k')
&=
g^0_k(k+t^h_i)-\gamma_k^h(k+t^h_i;k)
=
g^0_k(k+t^h_i)-g_k^h(k+t^h_i)
\end{align*}
and that
\begin{equation*}
\int_{k-h}^{k+h}\big[D^2\gamma_k^h(x;k)\big]^2dx
	=
\int_{k'-h}^{k'+h}\big[D^2\gamma_k^h(x+y;k')\big]^2dx
\end{equation*}
Using the notation of \eqref{I}, we conclude that
\begin{equation*}
I_k(h)
	=
I_k\big(h;\gamma^h_k(\cdot;k)\big)
	=
I_{k+y}\big(h;\gamma^h_k(\cdot;k')\big)
	\ge
I_{k'}(h)
\qquad k>0,y\in\R
\end{equation*}
The same inequality holds after exchanging $k$ for $k'$ so that $\gamma^h_k(\cdot;k')$ 
and $g^h_{k'}$ are both $\mathscr C^2$ splines of degree 3 solving \eqref{spline} 
and thus coincide, by uniqueness. We conclude that $g^h_{k+y}(x)=g_k^h(x-y)$
and thus, if $0\le a\le1$ and $k\le ak_1+(1-a)k_2$,
\begin{align*}
ag^h_{k_1}(x)+(1-a)g^h_{k_2}(x)
	&=
ag^h_{k}\big((x+(k_1-k)\big)+(1-a)g^h_{k}(x+k_2-k)\\
	&\ge
g^h_{k}\big(x+ak_1+(1-a)k_2-k)\big)\\
	&\ge
g^h_{k}(x)
\end{align*}
proving (\textit{i}). 
\end{proof}

\begin{proof}[\textbf{Proof of Theorem \ref{th convergence}}]
Write
\begin{equation}
\label{qe}
q^\delta(k)
	=
F(k)
	+
\sum_{\tau\in Z}w_k^\delta(\tau)[F(\tau)-F(k)]
	+
\sum_{\tau\in Z}w_k^\delta(\tau)\varepsilon_\tau
	=
F(k)
	+
A_k^\delta(Z)
	+
\sum_{\tau\in Z}w_k^\delta(\tau)\varepsilon_\tau
\end{equation}
with $A^\delta_k(Z)=\sum_{\tau\in Z}w_k^\delta(\tau)[F(\tau)-F(k)]$. Then, 
by \eqref{ass a}
\begin{align*}
P_Z\Big(\big(q^\delta(k)-F(k)\big)^2\Big)
	&=
\sum_{\sigma,\tau\in Z}w_k^\delta(\sigma)w_k^\delta(\tau)
P_Z\Big(\big(F(\sigma)-F(k)+\varepsilon_\sigma\big)
\big(F(\tau)-F(k)+\varepsilon_\tau\big)\Big)\\
	&=
A_k^\delta(Z)^2
	+
\big[P_Z(\varepsilon_\sigma^2)-P_Z(\varepsilon_\sigma\varepsilon_\tau)\big]
\sum_{\tau\in Z}w_k^\delta(\tau)^2
	+
P_Z(\varepsilon_\sigma\varepsilon_\tau)
	+
2P_Z(\varepsilon_\tau)A_k^\delta(Z)
\end{align*}
By \eqref{w>0,i>0} we know that $w_k^\delta(\tau)=0$ when $\abs{k-\tau}>(1+\delta)M_Z$ 
and $\tau>0$. Using Taylor expansion with Lagrange remainder we get
\begin{align*}
\babs{A_k^\delta(Z)}
	&\le
w_k^\delta(0)[F(0)-F(k)]
+\sum_{\tau\in Z\setminus\{0\}}w_k^\delta(\tau)\Big[DF(k)\abs{\tau-k}+\frac{(\tau-k)^2}{2}D^2F(k)
+\frac{\abs{\tau-k}^3}{3!}D^3F(x_\tau)\Big]\\
	&\le
F(0)\sset{\tau_1>k-\delta M_Z}+\abs{DF(k)}(\delta+1) M_Z+D^2F(k)(\delta+1)^2M_Z^2
+\frac{\alpha}{3!}(1+\delta)^3M_Z^3
\end{align*}
where we made use of \eqref{w>0,i=0}. 
Under \eqref{ass c} this implies
\begin{align*}
\lim_{M\to0}E\Big(A_k^\delta(Z)^2+2P_Z(\varepsilon_\tau)A_k^\delta(Z)\Big)
	=
\lim_{M\to0}E\Big(\big(A_k^\delta(Z)^2+2P_Z(\varepsilon_\tau)A_k^\delta(Z)\big)
\sset{\tau_1\le k-\delta M_Z}\Big)
	=
0
\end{align*}•
If $i\ge1$ we have 
$w_k^\delta(\tau_i)=D_{i+1}(g_k^\delta)-D_i(g_k^\delta)$ (with $\tau_{I+1}=\tau_I+1$, 
conventionally) so that 
\begin{align*}
w_k^\delta(\tau_i)
	&=
D_i(g_k^\delta)-D_{i-1}(g_k^\delta)\\
	&\le
Dg_k^\delta(\tau_{i+1})-Dg_k^\delta(\tau_{i-1})&\text{(convexity)}\\
	&=
D^2g_k^\delta(k)(\tau_{i+1}-\tau_{i-1})+\frac{(\tau_{i+1}-k)^2}{2}D^3g_k^\delta(x_{\tau_{i+1}})-\frac{(k-\tau_{i-1})^2}{2}D^3g_k^\delta(x_{\tau_{i-1}})\\
	&\le
2D^2g_k^\delta(k)M_Z+\alpha[1+(1+\delta)^2]M_Z^2
\end{align*}•
and thus 
\begin{align}
\label{w le}
\sum_{\tau\in Z}w_k^\delta(\tau)^2
	\le 
\sset{\tau_1>k-\delta M_Z}+2D^2g_k^\delta(k)M_Z+\alpha[1+(1+\delta)^2]M_Z^2
\end{align}
The claim follows from
\begin{align*}
\lim_{M\to0}E\Big((q^\delta(k)-F(k))^2\Big)
	&=
\lim_{M\to0}E\Big(
\big[P_Z(\varepsilon_\sigma^2)-P_Z(\varepsilon_\sigma\varepsilon_\tau)\big]
\sum_{\tau\in Z}w_k^\delta(\tau)^2
	+
P_Z(\varepsilon_\sigma\varepsilon_\tau)
\Big)\\
	&\le
\lim_{M\to0}E\Big(
2\sup_{\sigma,\tau}P_Z(\varepsilon_\sigma\varepsilon_\tau)
\sum_{\tau\in Z}w_k^\delta(\tau)^2\sset{\tau_1\le k-\delta M_Z}
\Big)+E(\varepsilon_\sigma\varepsilon_\tau)\\
	&=
0
\end{align*}
\end{proof}


\begin{thebibliography}{99} 

\bibitem{ait sahalia duarte} 
Y. A\"it-Sahalia, J. Duarte (2003), 
\textit{Nonparametric Option Pricing Under Shape Restrictions}, 
J. Econometrics \textbf{116}, 9-47.

\bibitem{ait sahalia lo} 
Y. A\"it-Sahalia, A. Lo (1998), 
\textit{Nonparametric Estimation of State-Price-Densities Implicit in Financial Asset Prices}, 
J. Finance \textbf{53}, 499-547.

\bibitem{banz miller} 
R. W. Banz, M. H. Miller (1978), 
\textit{Prices for State-Contingent Claims: Some Estimates and Applications}, 
J. Business \textbf{51}, 653-672.

\bibitem{bollen}
N. P. B. Bollen, T. Smith, R. E. Whaley (2004), 
\textit{Modeling the bid/ask Spread: Measuring the Inventory-Holding Premium}, 
J. Financ. Econ. \textbf{72}, 97-141.

\bibitem{breeden litzenberger} 
D. Breeden, R. Litzenberger (1978), 
\textit{Prices of State-Contingent Claims Implicit in Option Prices}, 
J. Business \textbf{51}, 621-651.

\bibitem{ThOpt}
G. Cassese (2014), 
\textit{Asset Pricing in an Imperfect World}, 
mimeo, http://arxiv.org/abs/1410.6408.

\bibitem{das newey vella}
M. Das, W. K. Newey, F. Vella (2003), 
\textit{Estimation of Sample Selection Models},
Rev. Econ. Stud. \textbf{70}, 33-58.

\bibitem{dykstra}
R. L. Dykstra (1983), 
\textit{An Algorithm for Restricted Least Squares Regression}, 
J. Amer. Stat. Ass. \textbf{78}, 837-842.

\bibitem{eubank}
R. L. Eubank (1999), 
\textit{Nonparametric Regression and Spline Smoothing},
Marcel Dekker, New York - Basel.

\bibitem{fengler}
M. R. Fengler (2009), 
\textit{Arbitrage-Free Smoothing of the Implied Volatility Surface}, 
Quant. Finance \textbf{9}, 417-428.

\bibitem{fengler hin}
M. R. Fengler, L.-Y. Hin (2014),
\textit{Semi-nonparametric Estimation of the Call-Option Price Surface under 
Strike and Time-to-expiry No-arbitrage Constraints},
J. Econometrics \textbf{184}, 242-261.

\bibitem{gagliardini gourieroux renault}
P. Gagliardini, C. Gourieroux, E. Renault (2011) 
\textit{Efficient Derivative Pricing by the Extended Method of Moments}, 
Econometrica \textbf{79}, 1181-1232.

\bibitem{garcia ghysels renault}
R. Garcia, E. Ghysels, E. Renault (2010), 
\textit{The Econometrics of Options Pricing}, 
in Y. A\"it-Sahalia, L. P. Hansen (Eds.) \textit{Handbook of Fianncial Econometrics},
vol. 1, 479-552, Amsterdam North-Holland.

\bibitem{hasbrouck} 
J. Hasbrouck (2002), 
\textit{Stalking the \quot{Efficient Price} in Market Microstructure Specifications: 
an Overview}, 
J. Financial Markets \textbf{5}, 329-339.

\bibitem{heckman}
J. J. Heckman (1979), 
\textit{Sample Selection Bias as a Specification Error},
Econometrica \textbf{47}, 153-161.

\bibitem{huang stoll} 
R. D. Huang, H. R. Stoll (1997), 
\textit{The Components of the bid/ask Spread: a General Approach}, 
Rev. Financial Studies \textbf{10}, 995-1034.	

\bibitem{jackwerth rubinstein} 
J. C. Jackwerth, M. E. Rubinstein (1996), 
\textit{Recovering Probability Distributions from Option Prices}, 
J. Finance \textbf{51}, 1611-1631.

\bibitem{mammen}
E. Mammen, C. Thomas-Agnan (1999), 
\textit{Smoothing Splines and Shape Restrictions}, 
Scand. J. Statist. \textbf{26}, 239-252.

\bibitem{melick thomas}
W. R. Melick, C. P. Thomas (1997),
\textit{Recovering Asset's Implied PDF from Option Prices: An Application
to Crude Oil during the Gulf Crisis},
J. Financ. Quant. Analysis \textbf{32}, 91-115.

\bibitem{ritchey}
R. J. Ritchey (1990), 
\textit{Call Option Valuation for Discrete Normal Mixtures},
J. Financ. Res. \textbf{13}, 285-296.

\bibitem{rompolis tzavalis}
L. Rompolis, E. Tzavalis (2008),
\textit{Recovering Risk Neutral Densities from Option Prices: A New Approach}, 
J. Financ. Quant. Analysis \textbf{43}, 1037-1053.

\bibitem{soderlind}
P. S\"oderlind (2000),
\textit{Market Expectations in the UK before and after the ERM Crisis},
Economica \textbf{67}, 1-18.

\bibitem{soderlind svensson}
P. S\"oderlind, L. Svensson (1997),
\textit{New Techniques to Extract Market Expectations from Financial Instruments},
J. Monetary Econ. \textbf{40}, 383-429.

\bibitem{turlach}
B. A. Turlach (2005), 
\textit{Shape Constrained Smoothing Using Smoothing Splines},
Comp. Stat. \textbf{20}, 81-103.

\bibitem{yatchew}
A. Yatchew (1998), 
\textit{Nonparametric Regression Techniques in Economics}, 
J. Econ. Lit. \textbf{36}, 669-721.

\bibitem{yatchew hardle} 
A. Yatchew, W. H\"ardle (2006), 
\textit{Nonparametric State Price Density Estimator Using Constrained 
Least Squares and Bootstrap}, 
J. Econometrics \textbf{133}, 579-599.

\bibitem{yin wang qi}
H. Yin, Y. Wang, L. Qi (2009), 
\textit{Shape-Preserving Interpolation and Smoothing for Options Market Implied Volatility}, 
J. Optim. Theory Appl. \textbf{142}, 243-266.

\end{thebibliography}
\end{document}